\documentclass[traditabstract]{aa} 
\usepackage{graphicx}
\usepackage{txfonts}
 \usepackage[T1]{fontenc}
\usepackage{color}
\usepackage[]{natbib}

	\newcommand{\arc}{$^{\prime\prime}$}
	\newcommand{\msun}{\ensuremath{\mathrm{M}_{\odot}}}
	\newcommand{\lsun}{L$_{\odot}$}
	\newcommand{\sfr}{M$_{\odot}$ yr$^{-1}$}

\begin{document}
   \title{Strongly star-forming rotating disks in a complex merging system at $z=4.7$ as revealed by ALMA}

   \author{S. Carniani
          \inst{1,2},
          A. Marconi
          \inst{1,2},
          A. Biggs 
          \inst{3},
          G. Cresci
          \inst{4},
          G. Cupani
          \inst{5},
          V. D'Odorico
          \inst{5},
          E. Humphreys 
          \inst{3},
          R. Maiolino 
          \inst{6,7},
          F. Mannucci
          \inst{2},
         P. Molaro
          \inst{5},
            T. Nagao
          \inst{8},         
          L. Testi 
          \inst{3},
          \and M. A. Zwaan
          \inst{3}
          }

   \institute{Dipartimento di Fisica e Astronomia, Universit\`a di Firenze, Via G. Sansone 1, I-50019, Sesto Fiorentino (Firenze), Italy
                      \and
             INAF - Osservatorio Astrofisico di Arcetri, Largo E. Fermi 5, I-50125, Firenze, Italy
          \and
             European Southern Observatory, Karl-Schwarzschild-Str. 2, 85748 Garching b. M\"unchen, Germany 
         \and
              INAF - Osservatorio Astronomico di Bologna, Via Ranzani 1, 40127 Bologna, Italy
          \and
                     INAF - Osservatorio Astronomico di Trieste, via Tiepolo 11, 34143 Trieste, Italy
         \and
             Cavendish Laboratory, University of Cambridge, 19 J. J. Thomson Ave., Cambridge CB3 0HE, UK
                      \and
               Kavli Institute for Cosmology, University of Cambridge, Madingley Road, Cambridge CB3 0HA, UK 
              \and
           Department of Astronomy, Kyoto University, Kitashirakawa-Oiwake-cho, Sakyo-ku, Kyoto 606-8502, Japan
             }

   \date{Received; accepted}

  \abstract{ 
  We performed a kinematical analysis of the [CII] line emission of the BR 1202-0725 system at $z\sim 4.7$ using ALMA Science Verification observations. The most prominent sources of this system are a quasar and a submillimeter galaxy, separated by  a projected distance of about $\sim$24 kpc and characterized by  very high star-formation rates, higher than $\sim 1000\,\mathrm{M_\odot\,yr^{-1}}$. However, the ALMA observations reveal that these galaxies apparently have undisturbed rotating disks, which is at variance with the commonly accepted scenario in which strong star formation activity is induced by a major merger.
We also detected faint components which, after spectral deblending, were spatially resolved from the main quasar (QSO) and submillimeter galaxy (SMG) emissions. The relative  velocities and positions of these components are compatible with orbital motions within the gravitational potentials generated by the QSO host galaxy and the SMG, suggesting that they are smaller galaxies in interaction or gas clouds in accretion flows of tidal streams. Moreover, we did not find any clear spectral evidence for outflows caused by AGN or stellar feedback.
This suggests that the high star formation rates might be induced by interactions or minor mergers with these companions, whichdo not affect the large-scale kinematics of the disks, however. Alternatively, the strong star formation may be fueled by the accretion of pristine gas from the host halo.
Our kinematical analysis also indicates that  the QSO and the SMG have similar dynamical masses, mostly in the form of molecular gas, and that the QSO host galaxy and the SMG are seen close to face-on with  slightly different disk inclinations: the QSO host galaxy is seen almost face-on ($i\sim15^\circ$), while the SMG is seen at higher inclinations  ($i\sim 25^\circ$).
Finally, the ratio between the black hole mass of the QSO, obtained from new XShooter spectroscopy, and the dynamical mass of the host galaxy is similar to value found in very massive local galaxies, suggesting that the evolution of black hole galaxy relations  is probably better studied with dynamical  than with stellar host galaxy masses.}

\authorrunning{Carniani et al.}
\titlerunning{Rotating disks in a strongly star-forming merging system at $z=4.7$}
   \maketitle

\section{Introduction}
The growth of galaxies is closely traced by  star formation, and the cosmological evolution of the global star formation rate (SFR) density is thus a fundamental element for understanding galaxy evolution from high to low redshift. The cosmic SFR density peaks at about z=1$\sim$3, as illustrated by the so-called Madau-Lilly diagram \citep{lilly:1996,madau:1996,haarsma:2000,hopkins:2001}. 
Until recently, galaxy evolutionary models usually associated this peak with the increased rate of galaxy mergers and interactions in that redshift range. In particular,  the merger process is at the base of the formation of  massive galaxies: the majority of the stars are formed following major-merger episodes that induce high SFRs (from several hundreds to over a thousand $M_{\odot} \, {\rm yr}^{-1}$; \citealt{di-matteo:2005,guyon:2006,sijacki:2011,valiante:2011}) 
But a different point of view has emerged very recently, according to which the peak of the cosmic SFR density is expected to follow the evolution of the cosmic gas inflow rate. This is suggested by the existence of a main-sequence  of star-forming galaxies at any redshift , which is most likely fueled by continuous gas accretion \citep{keres:2005,genel:2008,conselice:2013}.

Strongly starbursting outliers of such a main sequence do exist and represent an increasingly important phase at higher redshifts: these outliers are thought to be mostly merger-induced star-forming objects, such as submillimeter  and ultraluminous infrared galaxies  (\citealt{tacconi:2010}),  even if some authors have  found that the fraction of merger systems is  small \citep{rodighiero:2010}. 
Whatever the origin of the evolution of the cosmic SFR density, submillimeter galaxies (SMGs) appear to fit well within a merger-induced star formation scenario. They are dust-obscured strongly star-forming  galaxies  whose number density peaks around z$\sim$2  \citep{Chapman:2003,Chapman:2005}, although recent ALMA surveys have shown that the SMG redshift distribution probably peaks at even higher redshift (z$\sim$3.5,  \citealt{Weis:2013}). Their luminosities are in excess of 10$^{12} \, L_{\odot}$, their SFRs are of about $\sim 10^2-10^3 M_{\odot} \, {\rm yr}^{-1}$, and their stellar masses are around $10^{11} M_{\odot}$, although recent estimates made by different authors  vary  by a factor $\sim 6$ \citep{magnelli:2010,chapman:2010,hainline:2011,michaowski:2010,magnelli:2012}. When combining these properties with typical gas fractions of 40$\%$ \citep{narayanan:2012}, it is  natural to assume that SMGs are the late stages of major-merger events.

However, in contrast with the  major-merger scenario, a large portion of the strongly star-forming galaxies at z$\sim$1-3, especially among Lyman-break galaxies, does not show the expected   disturbed kinematics  but is characterized by regularly rotating  disks \citep{forster-schreiber:2009,epinat:2009,cresci:2009,gnerucci:2011a}.  This observational result has prompted the suggestion that even stronger star formation may  be fueled by the accretion of pristine gas from the halo and dynamical instabilities within the massive gaseous disks \citep{genel:2008,dekel:2009,bouche:2010,cresci:2010}. Another evidence of this smooth accretion is provided by the so-called fundamental metallicity relation, a tight relation for star-forming galaxies that connects gas metallicity, SFR and stellar mass, and which does not evolve with redshift up to $z\sim 3$ \citep{mannucci:2010,Dayal:2013, bothwell:2013}. 

Star formation is not the only ingredient in the formation of galaxies. The discovery of the correlations between the masses of supermassive black holes and the properties of their host spheroids \citep{ferrarese:2000,gebhardt:2000,marconi:2003,Gultekin:2009,mcconnell:2013} and the observed cosmological evolution of the AGN number density, which follows that of the  cosmic SFR density \citep{marconi:2004,merloni:2008a,shankar:2010a}, have suggested that the growth of black holes (BHs) is intimately linked with that of their host galaxies: this so-called co-evolution is also based on the merger paradigm (e.g. \citealt{hopkins:2007}). In particular,  in the most luminous AGN,  major-merger episodes  destabilize large amounts of gas, which forms stars and eventually accretes onto the central supermassive black holes. The luminous AGN that are then activated   have feedback onto their host galaxies: fast radiation-driven outflows start  sweeping away the surrounding gas,  quenching both black hole accretion and star formation. This is the beginning the observed black hole galaxy correlations and explains the decrease of the cosmic SFR and AGN densities below $z\sim1$. 
The co-evolution scenario then implies that there should be a population of active, clustered star-forming galaxies at high redshift ($z>3$). 

Observations of the morphology and kinematics of molecular gas in high-redshift galaxies may help to understand the main mechanism that caused the peak of the cosmic SFR density. In particular, the kinematics of gas in  merging systems might indicate whether star formation activity is indeed associated with disturbed kinematics.
 Moreover, the distribution of ionized and molecular gas  in active high-redshift sources such as quasars might reveal outflows, as expected from the feedback model.

BR 1202-0725 is  an ideal system to test these scenarios as it is composed of strongly star forming interacting galaxies at redshift z$\sim$4.7. The most prominent sources are a quasar  (QSO) at the southeast and a submillimeter galaxy (SMG) at the northwest. The bright QSO (M$_{\rm B}$=-28.5) presents broad Ly$\alpha$ emission extended toward the northwest, which suggests  interaction with the SMG \citep{Hu:1996}. The SMG has been observed for the first time at 1.4 GHz \citep{carilli:2002} and is located about $\sim$24 kpc from the QSO \citep{wagg:2012}.  The observed flux level at 850 $\mu$m \citep{iono:2006} places the SMG at the bright end of the submillimeter galaxy population ( S[850 $\mu$m] $>$ 5 ), which represents only 20-30$\% $ of the submm sources responsible for the 850 $\mu$m  extragalactic background radiation (e.g.  \citealt{Coppin:2006}).  Both sources have a high far-infrared (FIR, 40-500 $\mu$m) luminosity, L$_{\rm FIR} > 10^{13} $\lsun \ \citep{iono:2006,omont:1996,carilli:2002} that likely indicates very strong star formation activity (SFR $> 10^3$ \sfr). Assuming that warm dust is heated by star formation activity and re-emits in the FIR, it is possible to derive an SFR$\sim$4500 \sfr \ for the QSO and an SFR$\sim$2000 \sfr \ for the SMG \citep{iono:2006}. 

The system BR 1202-0725 has been recently observed with ALMA during the science demonstration phase \citep{wagg:2012,Carilli:2013} and with the IRAM PdB Interferometer as reported by \citet{salome:2012}.
These authors derived SFRs of about 1000 \sfr\ and molecular gas masses of $\sim 5\times 10^{10}\,M_\odot$.
Moreover, the system has been found to be more complex than previously thought. \cite{wagg:2012} and \cite{Carilli:2013} detected two fainter sources, one southwest of the quasar and one in between the QSO and the SMG, which are the [CII] counterparts of known Ly$\alpha$ emitters; furthermore, the CO line profile and position-velocity diagram by  \cite{salome:2012} indicate  a close companion also for the SMG.
Intriguingly, the CO  position-velocity diagrams for both the QSO and the SMG suggest that the two galaxies have regularly rotating disks.

Here we perform a kinematical analysis of the [CII] line emission in BR 1202-0725 from the  observations obtained with ALMA during the science demonstration phase.
We show that, at variance with a simple major-merger scenario, the high SFRs in both the QSO host galaxy and the SMG are associated with apparently regularly rotating disks, suggesting that
the high SFRs are probably induced by minor mergers, interactions, or gas inflows that  do not strongly affect the large-scale disk kinematics. We determine the dynamical masses of the two galaxies and study the  relation between black hole and host galaxy for the QSO.
Data reduction is described in sec.~\ref{sec:data} and results from the data analysis are presented in sec.~\ref{sec:results}. In sec.~\ref{sec:kinematics} we present our kinematical analysis and the dynamical mass measurements of the QSO host galaxy and SMG. We discuss our findings in sec.~\ref{sec:discussion} and  finally draw our conclusions in sec.~\ref{sec:conclusions}.

Throughout  this paper we adopt the  standard cosmological parameters  $H_0=70\ \mathrm{km\ s^{-1} Mpc^{-1}}$,  $\Omega_M=0.30$  and $\Omega_\Lambda=0.70$, according to which 1 arcsec at $z=4.7$ corresponds to a proper distance of 6.48 kpc.

\section{Observations and data reduction}\label{sec:data}
BR1202-0725 was observed at 0.9 mm (334 GHz) using  the Atacama Large Millimeter/Submillimeter Array, ALMA, with 18 12-m and 3 7-m diameter antennas on 2012 January 16; for this work we took advantage of all the available data. The total on-source observing time was about 25 minutes with a maximum baseline length $\sim$~280 m. The weather conditions were generally very good with a precipitable water vapor of 0.7 mm, and the continuum sensitivity of ALMA was one order of magnitude higher than previous submillimeter observations of the system. 

The receivers were tuned to a [CII] redshifted frequency of about 333 GHz (the rest frequency of [CII] is 1900.539 GHz). The ALMA correlator has a  total bandwidth of 7.5 GHz and was divided in to four spectral windows of $\sim 1.8$ GHz  with a channel width of 15.6 MHz ($\sim$ 14 km s$^{-1}$). The data were calibrated by using the CASA software. The amplitude and phase fluctuations  were calibrated by observations of 3C279, for which we took the flux of 15 Jy obtained from SMA observation at 850 $\mu$m. Afterwards we used  observations of Titan to rescale the flux,  assuming a total value of 2.5 Jy.  The quasar 3C 279 was also used to calibrate the time-dependent gain and the bandpass.  Overall, we  estimate a 15$\%$ uncertainty on the measured absolute fluxes.

The high sensitivity of ALMA has allowed a self-calibration of the BR1202-0725 data using the line-free channels: we  used a 3-min averaging time for the complex gain solutions in the amplitude and phase self-calibration.

The continuum image was extracted using  all the line-free channels in the  last three spectral windows; the image was then cleaned using the CASA task "CLEAN" with ROBUST=1.5, achieving an angular resolution of 0.8\arc$\times$0.7\arc with the beam oriented at P.A. = 8$^{\circ}$. The achieved sensitivity in the continuum image is 0.2 mJy (\textit{rms}/beam).

A UV plane model of the continuum emission was made from the line-free channels of the last three spectral windows and was subtracted from the first and second spectral windows after linear extrapolation. Following the same deconvolution procedure as for the continuum image, we  then obtained the final line data cube with the  first two spectral windows rebinned to channels of $\sim$28 km/s . The spatial scale is 0.15\arc\ per pixel ($\sim$ 0.97 kpc per pixel). The  rms noise was estimated to be 2.0 mJy per beam per channel,  from several line-free channels. 

The continuum and line maps obtained by us are similar to those provided with the data release. Continuum and line fluxes are consistent within the errors and the main difference consists in our  choice of pixel scale and ROBUST values, which improved the spatial resolution by a factor~$\sim~2$.

\section{Results}\label{sec:results}

\begin{table*}
\caption{Summary of the  physical properties of the BR1202-0725 system from the  literature and from this paper.}           
\label{tab:properties}      
\centering          
\begin{tabular}{l c c c }     
			 \hline
			 \\
 & \multicolumn{2}{c}{Sources}			\\
 \\
  &  BR1202 N	& BR1202 S  & References \\
			 \\
			 \hline
			 \\
 RA									&12:05:22.98		&12:05:23.13				&  \\
 Dec									&-07.42.29,56		&-07.42.32,68			& 	\\
Integrated continuum flux at 900 $\mu$m [mJy] 	& 15.8$\pm$2.4 		& 14.8$\pm$2.2 			&  \\
Distance from S source [kpc] 						& 23.5$\pm$0.3 		& 	- 					& \\
FWHM from Gaussian fitting of continuum  	& 0.84\arc$\times$0.75\arc	& 0.81\arc$\times$0.78\arc	&	\\
Integrated [CII] flux [Jy/km \ s$^{-1}$]			& 9.06				& 8.45					&  \\
FHWM of [CII] line	 [km\ s$^{-1}$]				& 500$\pm$	20		& 300$\pm$20			&	\\
Redshift  										& 4.6891			& 4.6943				 & \\
L$_{\rm FIR}$ [ \lsun] 							& 1.5$\times10^{13}$	& 1.8$\times10^{13}$		&\\
SFR from continuum [\sfr]						& 2600				& 3200					& \\
SFR from [CII] [\sfr]							& 6800$\pm$1000 & 5000$\pm$1000&	  \\
L$^\prime_{CO}\,\, [10^{10}\,\mathrm{ K\, km/s\, pc^2}]$					& $7.8\pm1.2$		& 	$5.3\pm0.8$				& 	1	\\
$\alpha_{CO}$									& $0.6\pm0.2$		&	$0.6\pm0.2$			&	2\\
Molecular mass [$10^{10} M_\odot$]     			&  $4.6\pm1.7$ 		& 	$3.2\pm1.2$ 			& 	\\
Disk inclination [$^\circ$] 				& 25$\pm15$ 		& $ 15\pm10$ 					&  \\
P.A. of line of nodes [$^\circ$]									& 180$\pm$5 		& 230$\pm$10 				&  \\
Scale radius  [kpc] 							& 2.8$\pm$0.2 		&  2.4$\pm$0.2 				&  \\
$\log_{10} (M_{dyn}/M_\odot)$				& 10.8$\pm0.6$	  & $10.6^{+0.8}_{-0.4}$			& \\
$\log_{10} (M_{BH}/M_\odot)$	 								& - 					  & 9.1$\pm0.3$ 			& \\
$M_{BH}$ [10$^{9}$ \msun] 					& - 					  & 1.5$_{-0.8}^{+1.5}$ 		& \\
$\Sigma_{SFR}$ [\msun /yr kpc$^{-2}$]			&	80				  &  180						&	\\
$\Sigma_{gas}$ [\msun\,  pc$^{-2}$]				&	1900		 &   1800					&		\\
Dark halo mass [\msun]			 & \multicolumn{2}{c}{$> 3.2\times10^{11}$}  & \\

			 \hline	 
			 
\end{tabular}	
\tablebib{
(1) -\citet{salome:2012}	; (2)-\citet{bolatto:2013}	
}
\end{table*}

   \begin{figure}
   \centering
\includegraphics[width=10cm]{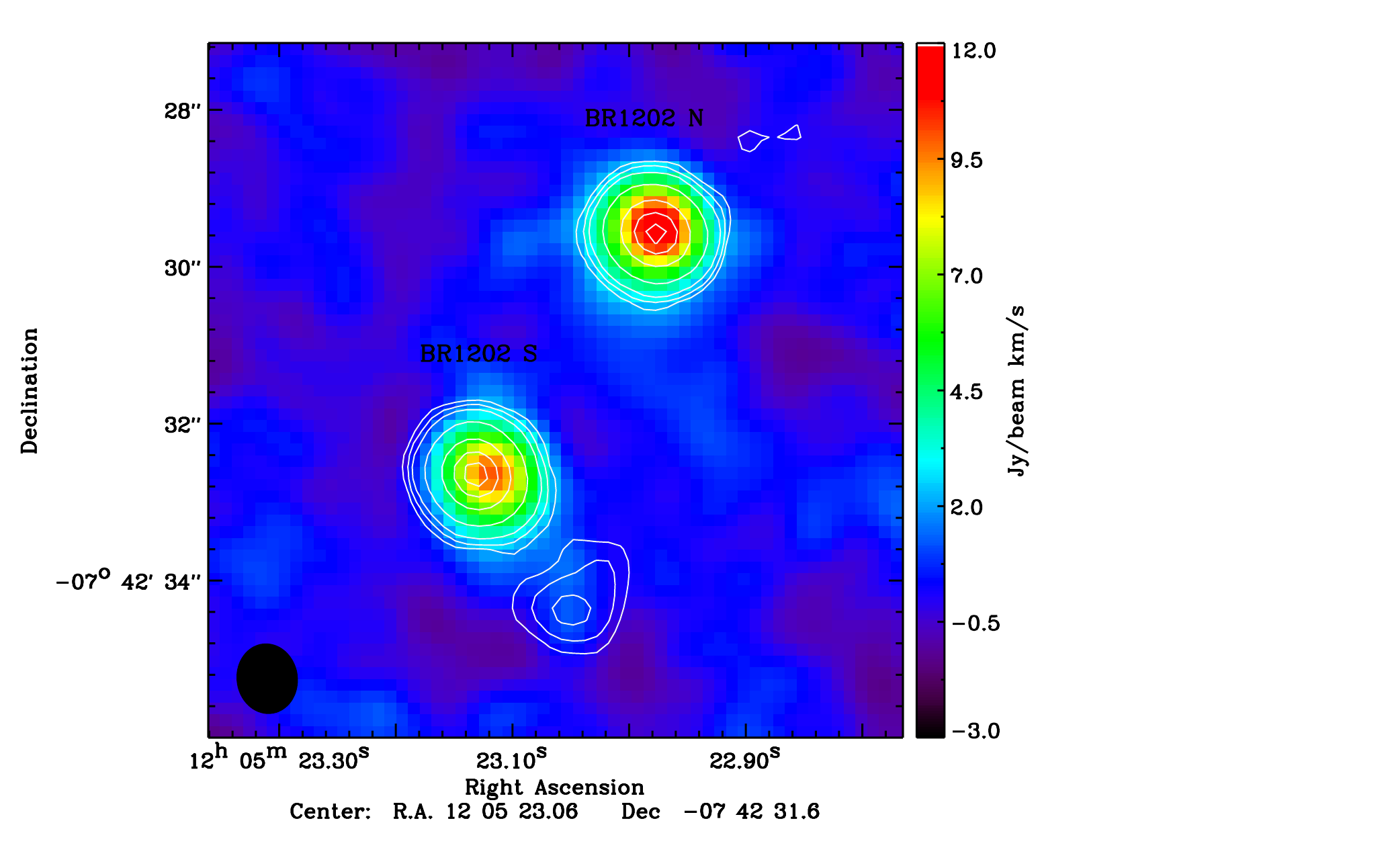}
      \caption{[CII] integrated intensity map in BR 1202-0725 with identification of the north and south sources. The rms noise is 0.6 Jy km s$^{-1}$. The synthesized beam (0.8\arc$\times$0.7\arc) is indicated by a filled black ellipse in the lower left corner of the plot. The continuum emission of BR1202-0725 a 334 GHz is overplotted as white contours. The contour levels are logarithmic [-3.4, -3.2, -3.0, -2.6, 2.2, 2.0, 1.9] of $\sigma$ with $\sigma$=0.2 mJy. The rms noise is 0.2 mJy beam$^{-1}$.
                    }
         \label{fig:continuum}
   \end{figure}

Table \ref{tab:properties} summarizes the physical properties of the BR1202-0275  system as measured in this paper or collected from the literature. Here we identified the two brightest sources with the labels N and S (north and south). The north source is the submillimeter galaxy (SMG) and the south source is the quasar (QSO) .

Figure \ref{fig:continuum} presents the [CII] intensity map of the BR1202-0275 system with the surface brightness contours of the 900 $\mu$m continuum emission overlaid. The integrated fluxes  of the N and S sources at 334 GHz and in the [CII] line and their centroid coordinates  (table  \ref{tab:properties})  are consistent with those found by \cite{iono:2006} and the previous analysis of these data by \cite{wagg:2012} and \cite{Carilli:2013}.  The projected distance between the two continuum sources is  23.5$\pm$0.3 kpc. The fainter continuum source discovered by \cite{wagg:2012} southwest of the QSO (Fig.~\ref{fig:continuum}) is detected  in the continuum at the $\sim$8$\sigma$ level ($\sigma$= 0.2 mJy beam$^{-1}$). This  source is not observed in the integrated  line  map of figure \ref{fig:continuum}  because of the large velocity range used in integration.

   \begin{figure*}
   \centering
\includegraphics[width=8cm]{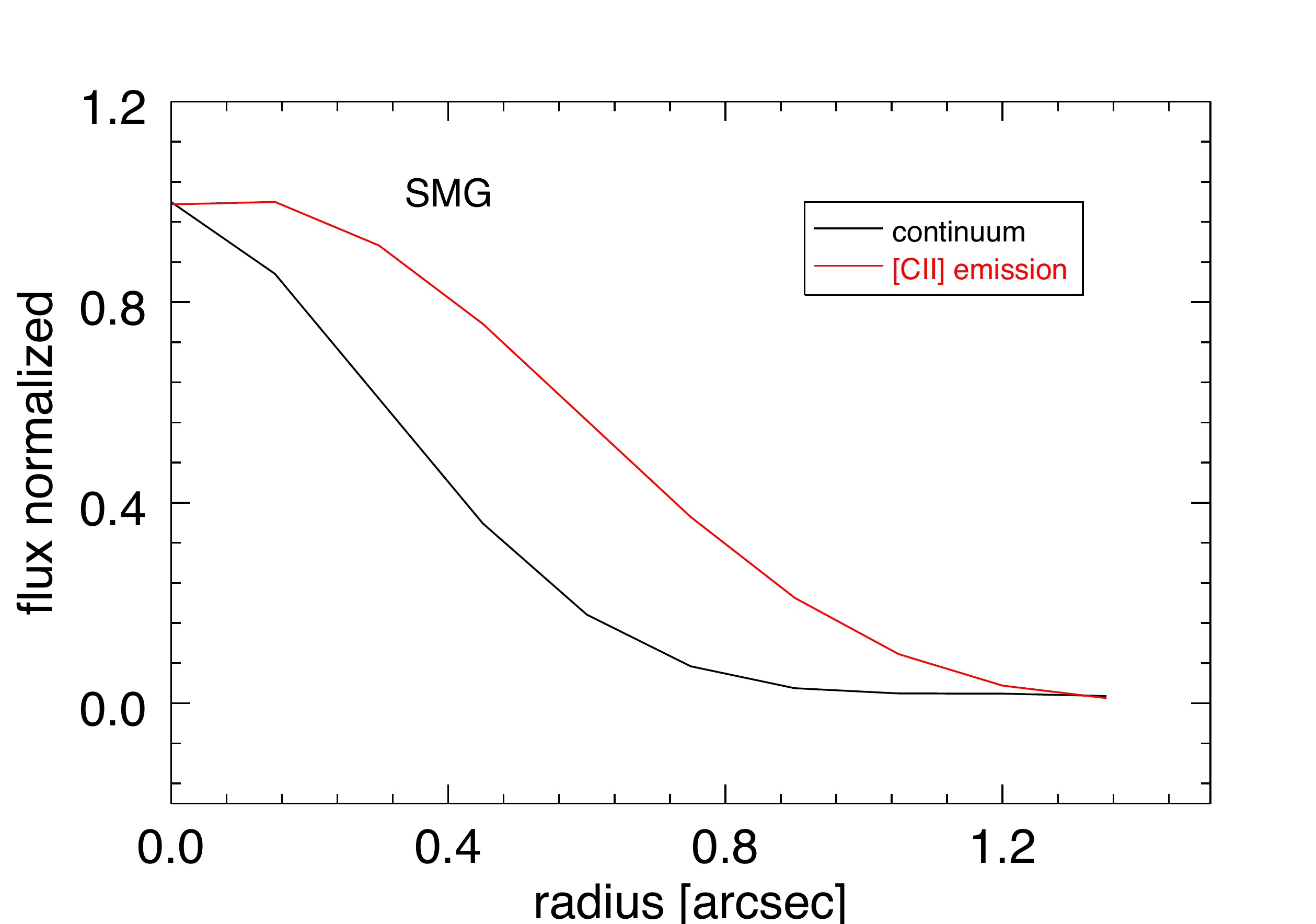}
\includegraphics[width=8cm]{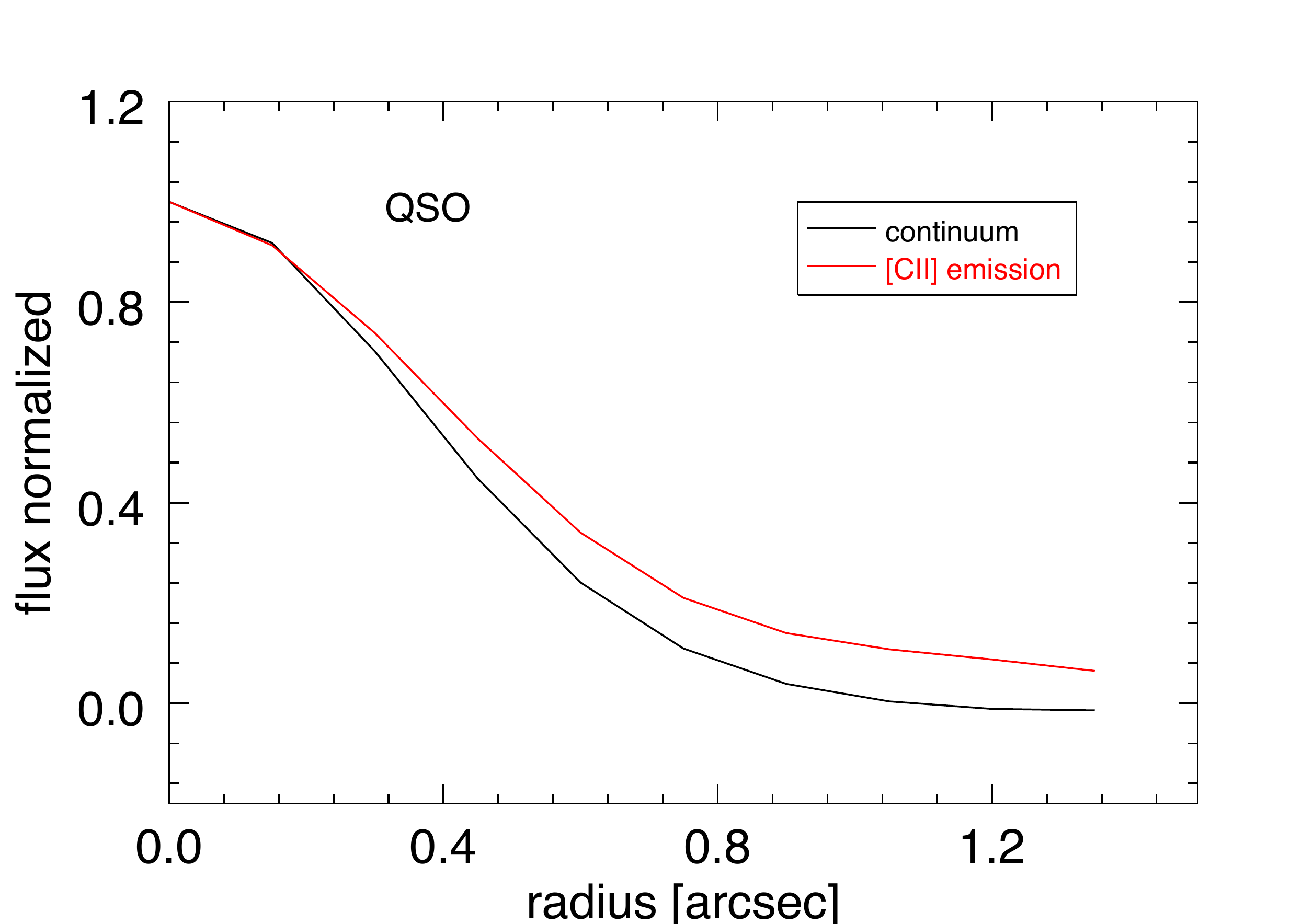}
      \caption{Radial surface brightness profiles of the submm galaxy (N source, left panel) and of the quasar (S source, right panel). The continuum and [CII] line emissions are denoted by the black and red lines, respectively.}
         \label{fig:line_profile}
   \end{figure*}

The continuum emission of the N and S sources is not spatially resolved because their FWHM are similar, within the errors, to that of the synthesized beam (Tab.~\ref{tab:properties}). The line emission of both sources, however, is spatially resolved compared to the continuum, as shown by the FWHM in table \ref{tab:properties} and by the radial surface brightness profiles presented in figure \ref{fig:line_profile}. As shown in the figure for both the N and S sources, the radially averaged  FWHM of  the [CII] emission along the x-axis (E-W)  is $\sim1.2$ times higher than that of the synthesized beam and of the continuum. 
Moreover, it is clear from figure  \ref{fig:continuum}  that the peak of [CII] emission in the QSO (S source) is shifted by 0.2\arc\ ($\sim 1.3$ kpc) with respect to the continuum. 

\begin{figure*}
 \centering
   \includegraphics[width=9cm]{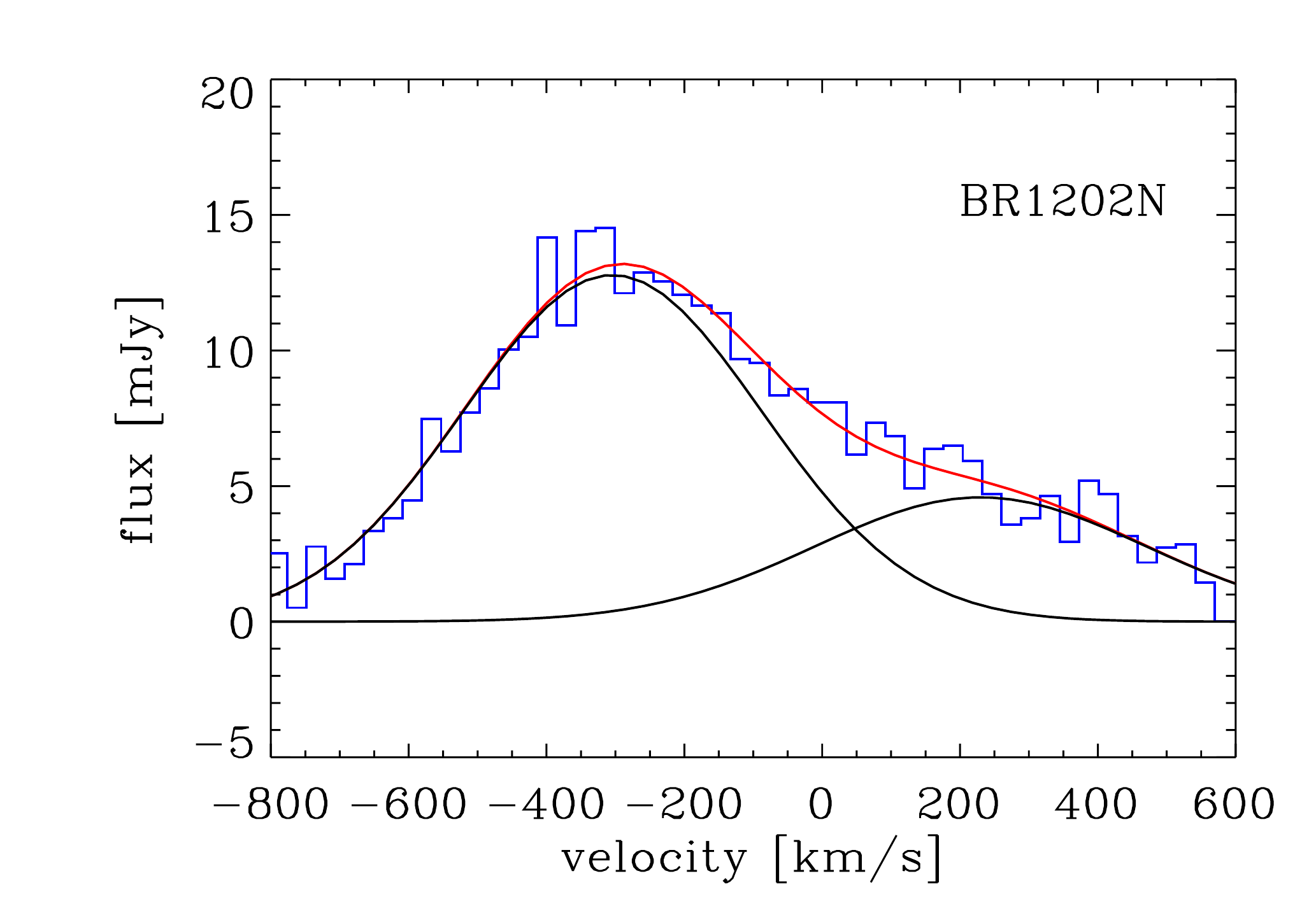}
  \includegraphics[width=9cm]{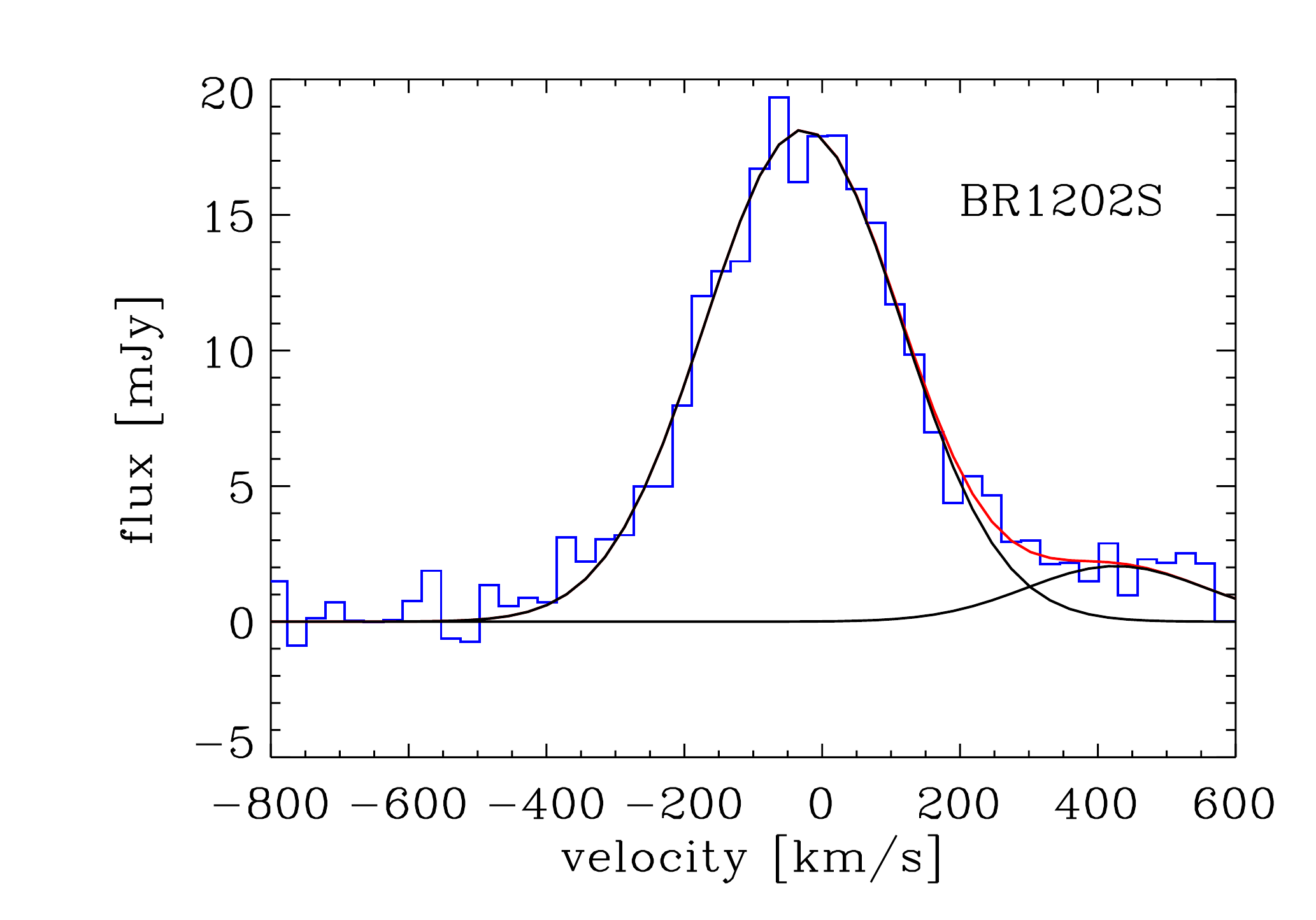}
   \caption{Integrated spectra of the N and S sources (left and right panels, respectively), targeting the [CII] line with a spectral resolution of 28 km s$^{-1}$ per channel. The spectra have been extracted  from the ALMA data cubes using beam-sized apertures centered on the sources.  All velocities are referred to  a redshift z=4.6949. The red curves denote the best fits, which are obtained with a double Gaussian function for both sources. For the QSO the width of the second component is fixed to that of the main component. For the SMG, the velocity of the fainter component is fixed at 250 km/s \citep{salome:2012}.}        
          \label{fig:spectrum}
    \end{figure*}

The integrated spectra of the N and S sources are shown in figure \ref{fig:spectrum}. 
The line profile of the QSO is generally symmetric, except for a ``red  wing'' at $\sim 500$ km/s that is probably associated with a faint companion or with an outflowing wind, as suggested by \cite{Carilli:2013}. To take into account the red wing, the spectrum was fit with two Gaussian functions with the same FWHM  because the red wing is at the edge of the band and not fully constrainable. As discussed below, this choice does not affect the kinematics of the QSO host galaxy. Conversely, the line profile of the SMG is asymmetric and broader than that of the quasar. It was fit with a combination of two Gaussians with the fainter component fixed at 250 km/s, as suggested by \cite{salome:2012} and explained in more detail below. The results, including average velocities  and velocity dispersions, are presented in table \ref{tab:properties}.
The redshifts of the two sources are determined from a Gaussian fit to the  [CII] lines and corresponds to z$= 4.6891\pm0.0004$ for the SMG and  z$=4.6943\pm0.0005$ for the QSO.
 In all cases [CII] velocities are referred to the systemic redshift of z=4.6949 adopted by \cite{salome:2012}; our velocity and the velocity dispersion measurements for [CII] are consistent with the measurements by  \cite{Carilli:2013}, \cite{wagg:2012}, \cite{iono:2006}, and with the corresponding CO values by \cite{salome:2012}.

\subsection{ Star formation rate and gas mass} 
We first estimated the SFR from the FIR continuum emission. To this aim, we fit the spectral energy  distributions (SEDs) of the two sources in the FIR-radio range using the model of Yun \& Carilli (2002) and assuming  typical values  for  the dust temperatures (T$_d$) and opacities (k$_\nu$) of SMGs and quasars (T$_d$ = 65 K for the SMG and T$_d$ = 50 K for the QSO; k$_\nu \sim \nu ^{1.5} $ ; see \citealt{iono:2006,beelen:2006,kovacs:2006} for more details). Fluxes for BR 1202N and BR 1202S were compiled from \citet[1.4 GHz]{carilli:2002}, \citet[4.8  GHz]{omont:1996},  \citet[1000 GHz]{benford:1999} and from our own ALMA measurement (334 GHz). The adopted SED model  is composed of thermal bremsstrahlung, nonthermal synchrotron and dust continuum emission; the latter  depends on the SFR, which is the only free parameter of  our SED fit. From this we estimate a star formation ratio (SFR) of $\sim 4000$ \sfr\ for the N galaxy and $\sim 5000$ \sfr\ for the S source.

Using the SED fitting only for estimating the FIR luminosity and applying the well-known SFR-FIR relation
\begin{equation}
SFR = 1.73\times 10^{-10} (L_{FIR}/L_\odot)\, M_\odot/yr
\end{equation}
by \cite{kennicutt:1998}, one obtains $\sim 2600$ and $\sim 3200$ \sfr\ for the SMG and QSO, respectively.  These values are similar to the previous ones and to those found by \cite{iono:2006} with the same relation.
The adopted SFR-FIR relation was calibrated with a Salpeter IMF; the use of a top-heavy Kroupa IMF would lower  the SFRs by a factor $\sim 1.6$ (see \citealt{salome:2012} and \citealt{mor:2012} for more details).

The SFR can also be estimated from the luminosity of the [CII]  emission line, since it has been shown to be a tracer of star formation activity in  different types of galaxies \citep{genzel:2000, malhotra:2001, luhman:2003}. Using the calibration of \cite{maiolino:2005}, we estimate an SFR of $\sim 6800$ \sfr\ for the SMG and  of $\sim 5000$ \sfr\ for the QSO.  The calibration of \cite{sargsyan:2012} would provide much lower SFRs of less than 900 \sfr\ for both sources, but this calibration may not be applicable since it is obtained from galaxies at lower redshifts ($z\sim 1-2$) and with lower star formation rates (SFR $< 500$ \sfr\ ).  Given the uncertainties on the use of [CII] as an SFR indicator and the possibility that [CII] could trace the cold neutral interstellar medium, in the following  we  used only SFRs based on the FIR luminosity and considered the values based on [CII] just as a confirmation of the very high SFR.  

We estimated the molecular gas masses  from $L^\prime_{\rm CO}$(5-4)  \citep{salome:2012} using the conversion factor 
$\alpha_{\rm CO} =0.6\pm 0.2 \ {\rm M_{\odot} \ ( \ K \ km/s \ pc^2)^{-1}}$ derived by \cite{papadopoulos:2012} from a large sample of starburst galaxies. For comparison, \cite{salome:2012} used the classical  value of $\alpha_{\rm CO} =0.8. \ {\rm M_{\odot} \ ( \ K \ km/s \ pc^2)^{-1}}$ from \cite{downes:1998}. As discussed in detail by  \cite{bolatto:2013}, the 
$\alpha_{\rm CO}$  conversion factor is still very uncertain especially at high redshift and, for this reason, 
 we adopted a factor 2 (0.3 dex) systematic uncertainty on $\alpha_{\rm CO}$.
The molecular masses are therefore $M_{CO} = (4.6\pm 1.7)\times 10^{10}$\msun\ and $M_{CO} = (3.2\pm 1.2)\times 10^{10}$\msun\ for the SMG and the QSO host galaxy, respectively. These measurements are also affected by the factor 2 systematic uncertainty on $\alpha_{\rm CO}$.

Summarizing,  the FIR and [CII] based indicators provide SFRs between $\sim 1000-5000$ \sfr\ for the two sources, which are associated with very large masses of molecular gas ($M_{mol}\sim 4\times 10^{10}\,M_\odot$). Overall, these results suggest extreme star formation activity, like that associated with major-merger events, and 
place the two sources at the higher end of the SFR distribution for star-forming galaxies. To verify this scenario and determine whether the two galaxies show the disturbed kinematics typical of a major merging event, we performed a detailed kinematical analysis.

\section {Kinematical analysis}\label{sec:kinematics}

In this section we perform a detailed kinematical analysis of the N and S sources, trying to assess whether they show the disturbances typical of a major-merger system, or are regularly rotating disks.
 
The QSO and the SMG probably identify two massive galaxies  that undergo an interaction or are in the early stages of a major-merger It is therefore possible to estimate the total mass of the system using a simple virial relation such as 
\begin{equation} 
M_{\rm tot}= d\,\Delta V ^2 /G = 3.2\times10^{11} \msun ,
\end{equation}
 where $d$ is the observed projected distance ($d=23.5$ kpc) and $\Delta V$ is the relative velocity ($\Delta V=240$ km/s). The dynamical mass was estimated assuming that the binary system is observed edge-on and that the galaxies are moving along the line of sight. This estimate is therefore a lower limit for the mass contained  within the  volume delimited by the location of two sources;  obviously, this is also the lower limit of the mass of the dark matter halo hosting the system.

The extended [CII] emission detected in the two sources (Fig.~\ref{fig:line_profile}) makes a spatially resolved kinematical analysis, which is now described for each source separately.
Following \cite{salome:2012}, we first analyzed the position-velocity (PV) diagrams of the [CII] line and  performed a full modeling of the observed kinematics.

 \begin{figure*}
 \centering
   \includegraphics[width=9cm]{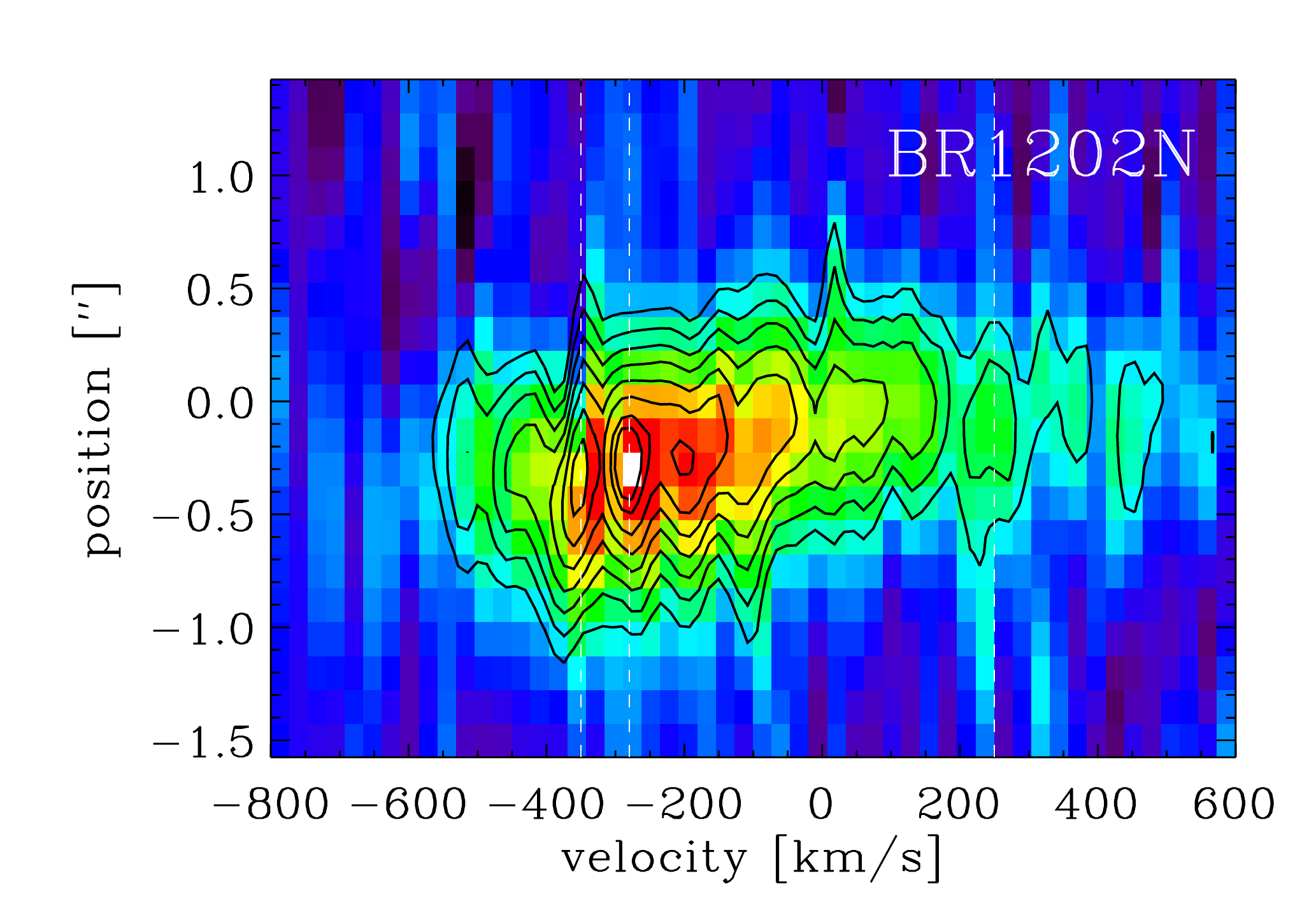}
  \includegraphics[width=9cm]{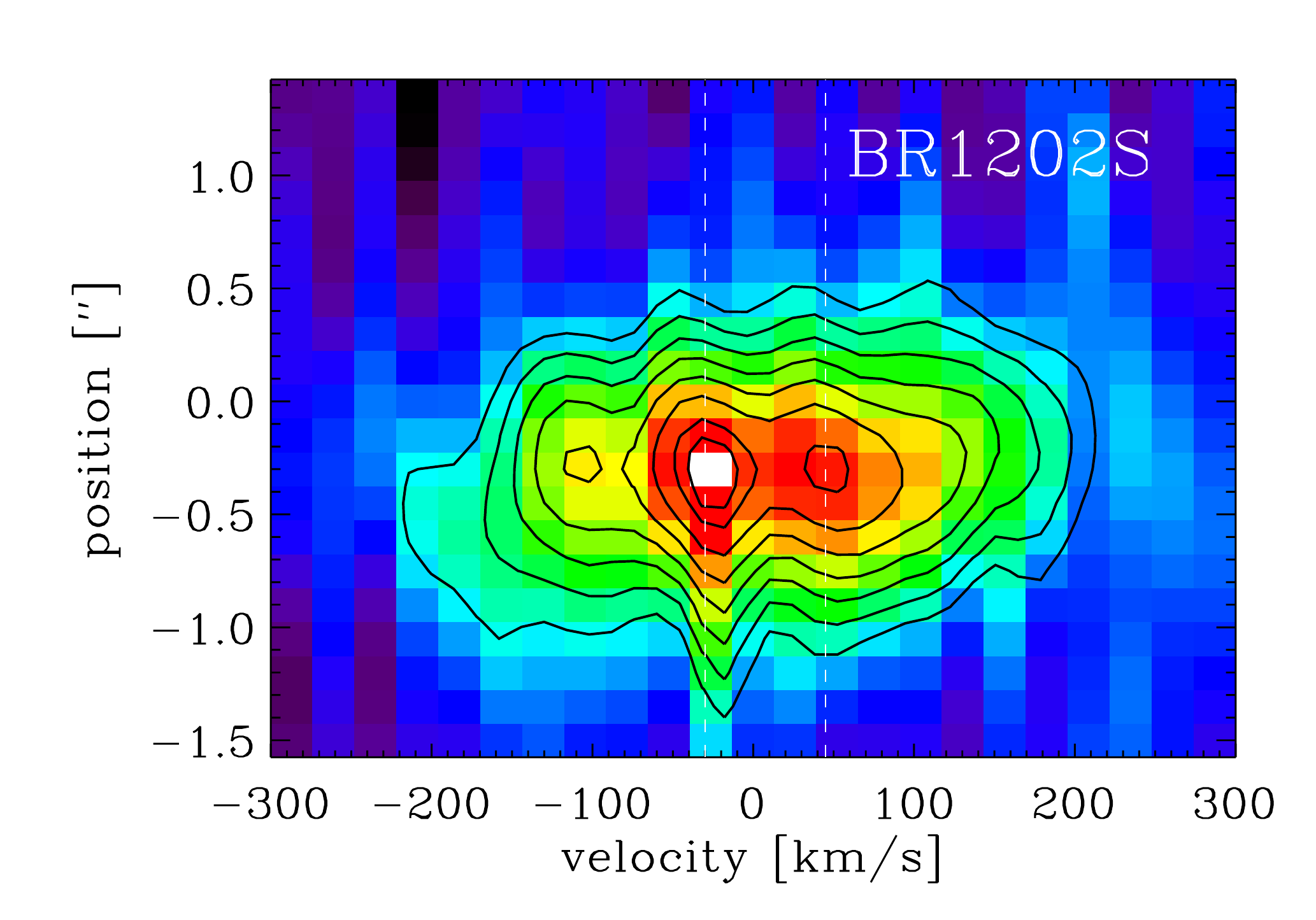}
   \caption{Left panel: the position-velocity diagram extracted from a three-pixel-wide E-W slit centered on the N source (SMG). The right panel is the  PV diagram  for a slit oriented N-S and centered on the  S source (QSO). The white dashed lines identify the components discussed in the text.}
              \label{fig:position_velocity}
    \end{figure*}
 \begin{figure*}
 \centering
   \includegraphics[width=8cm]{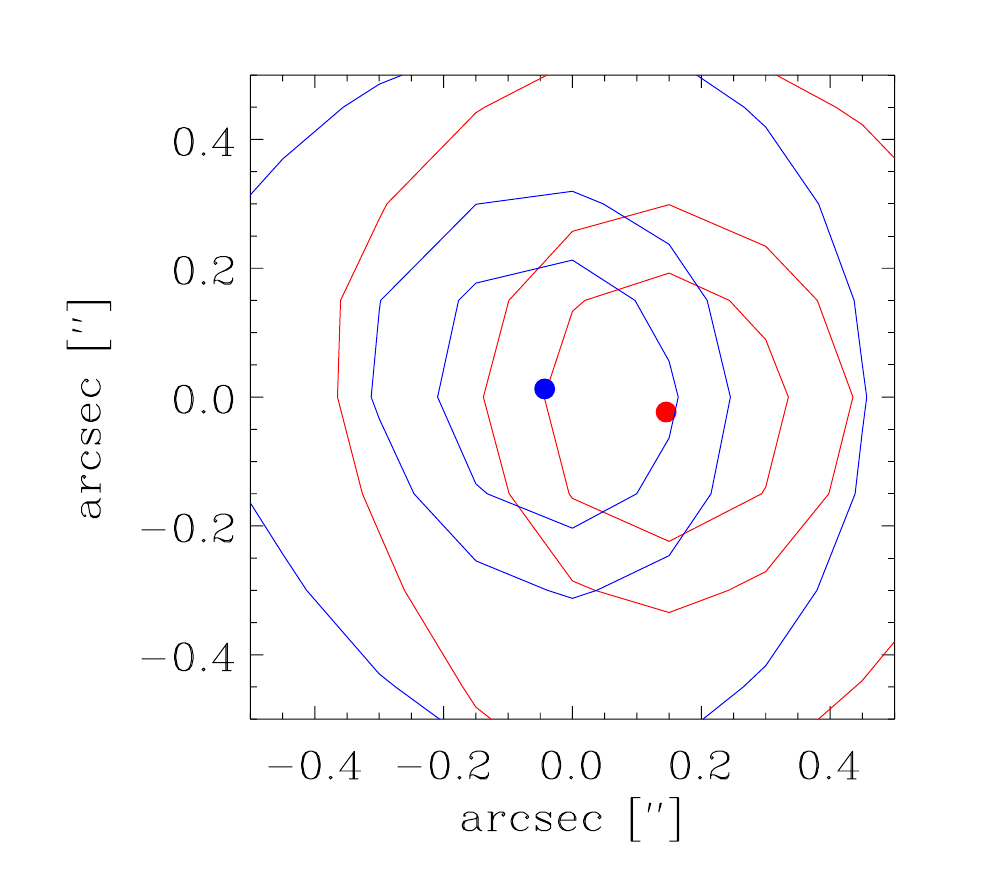}
  \includegraphics[width=8.1cm]{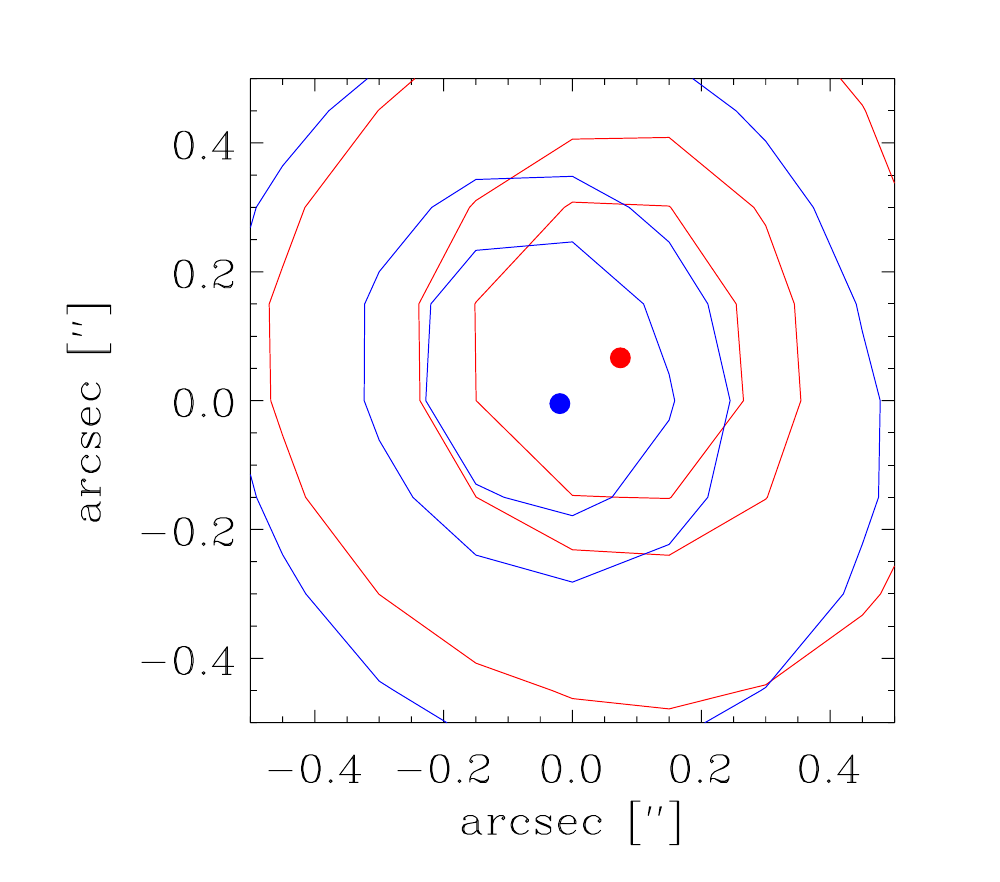}
   \caption{Surface brightness contours obtained by collapsing the red and blue channels in the submm galaxy (N source, left panel) and QSO (S source, right panel). Contours are at 90\%, 80\%, and 50\% of the peak values. 
  }
              \label{fig:spectroastrometry}
    \end{figure*}

\subsection{The submillimeter galaxy  BR1202 North}
   \begin{figure*}
   \centering
\includegraphics[width=0.8\linewidth]{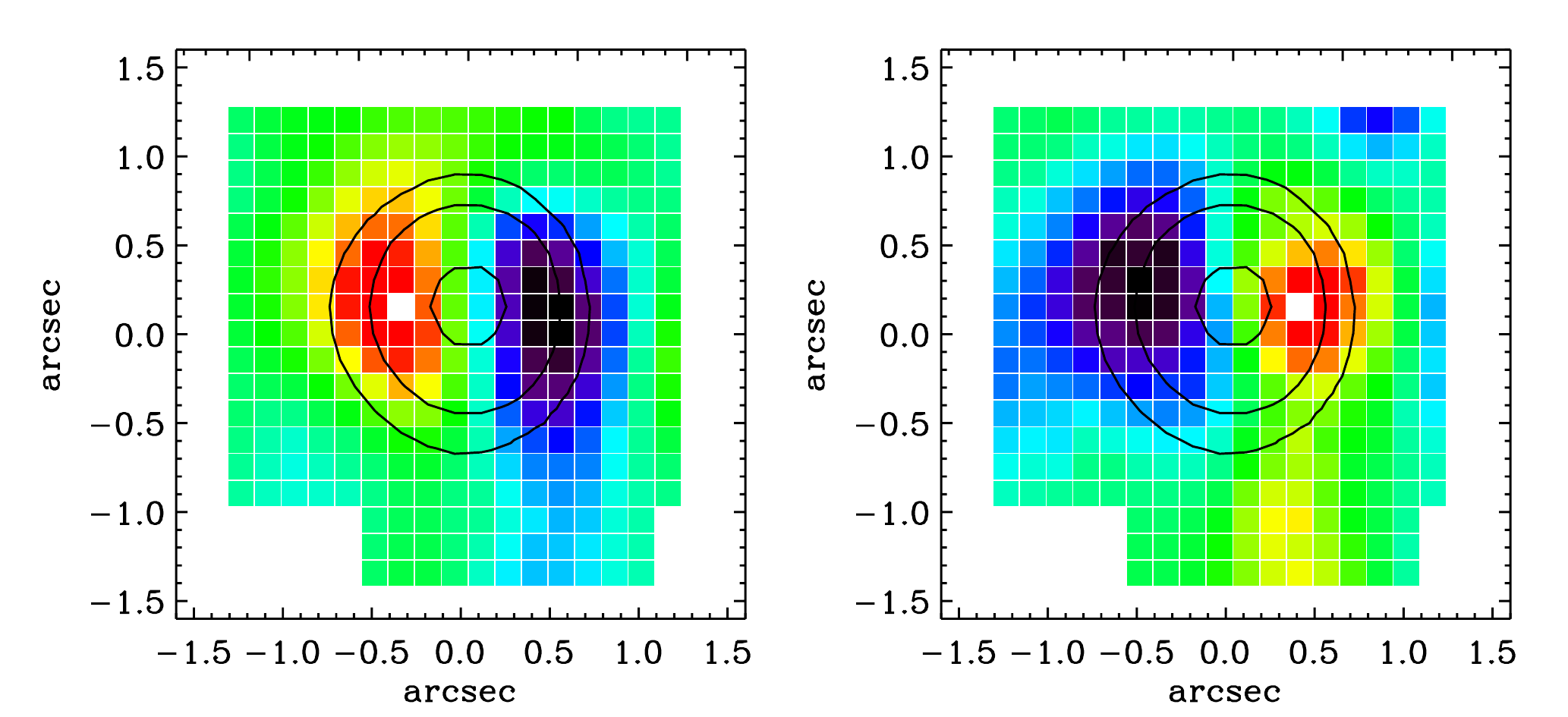}
      \caption{Blue and red residual maps of the submillimeter galaxy obtained assuming spatially unresolved kinematics (left and right panels, respectively).  Contours   represent the total line surface brightness at 90\%, 50\%, and 30\% of the peak value.
              }
         \label{fig:residual_map}
   \end{figure*}

The left panel of Figure \ref{fig:position_velocity} shows the [CII] position-velocity diagram for the submillimeter galaxy.
It has been obtained by extracting the spectrum from a three-pixel-wide (0.45\arc) slit centered on the [CII] surface brightness peak and oriented E-W, as in \cite{salome:2012}. The absolute velocity scale is referred to z = 4.6949.
This diagram presents at least three main [CII] components  at $\sim$-350 km s$^{-1}$,  $\sim$-280 km s$^{-1}$, and $\sim$250 km s$^{-1}$. As suggested by \cite{salome:2012} based on the CO PV diagram, the component at about $\sim$250 km s$^{-1}$ probably  indicates  another weak galaxy east of the SMG, which might be undergoing an interaction or a merger. This component is responsible for the asymmetric line profile of the SMG compared with that of the QSO (Fig.~\ref{fig:spectrum}).
The two main components are separated by $\Delta$v=70 km s$^{-1}$, resulting in a double-peaked line profile which, in some cases, can be the signature of a rotating disk. These separate peaks are not seen in the CO PV diagram by \cite{salome:2012}.

Before performing a spatially resolved kinematical analysis, we tested whether the kinematics is indeed spatially resolved and consistent with a rotating disk. 

The first test was to perform a spectroastrometric analysis by collapsing all the velocity channels blueward of the line peak and the redward ones in two separate images. If the SMG has a rotating disk and the rotation is resolved, the blue and red channel maps should have offset surface brightness centroids since they map the approaching and receding halves of the disk,  respectively. This is indeed shown in the left panel of fig.~\ref{fig:spectroastrometry} where the surface brightness peaks of the blue and red maps are separated  by $\sim 0.2$\arc\ (1.3 pixels corresponding to 1.2 kpc) along the E-W direction. Following \citep{gnerucci:2011}, we can therefore use spectroastrometry to obtain a first estimate of the dynamical mass of the SMG as 
\begin{equation}
M_{\rm dyn} \sin^2 i = (2.3\pm0.2)\times 10^{9} {\rm M}_{\odot} \left(\frac{FWHM}{100 \ {\rm km\, s^{-1}}} \right)^2 \left (\frac{r}{1 {\rm kpc}}\right)
\end{equation}
where $i$ is the disk inclination with respect to the line of sight, $FWHM$ is the full width at half maximum of the [CII] emission line emission, and $r$ is the half distance between the "red" and "blue" luminosity centroids. On top of the measurement errors, one should consider a 10\% systematic $rms$ error, deriving from the calibration of \cite{gnerucci:2011}, which was based on a sample of $z\sim 2-3$ Lyman-break galaxies with spatially resolved kinematics. Using $FWHM = 500\pm40 $ km/s, $r =0.6\pm0.2$ kpc, the spectroastrometric mass estimate is then $M_{\rm dyn} \sin^2 i =(3.5\pm0.4)\times10^{10}$\msun, with a systematic error of $10\% \,rms$. 
 We stress that this value  is valid only under the assumption of a virialized rotating disk. Indeed,  the results from the spectroastrometric analysis are consistent with the expectations from a virialized rotating disk, but it could  also be possible to reproduce them with outflowing gas in a particular configuration and therefore the more accurate analysis  described below is required. 

The second test consists of analyzing velocity residual map obtained with a pixel-by-pixel kinematical fitting after assuming that the sources are spatially unresolved. In this case, the line profile of the integrated [CII] emission plotted in Fig.~\ref{fig:line_profile} is expected to be equal to any line profile regardless of the spatial pixel from where they have been extracted. We therefore performed a pixel-by-pixel fit of the emission line with two Gaussian functions, only allowing for a variation of the total line flux, as expected from unresolved emission. We then obtained the residuals maps by collapsing the blue and red spectral channels, as we did for spectroastrometry.
The results of this exercise are shown in figure \ref{fig:residual_map}, where we show the blue and red residuals (left and right panels, respectively) with the contours of the total line surface brightness superimposed. The structure of the residuals is clearly what we would expect for a rotating disk after a line fit that does not allow for a variation of the line average velocity. Indeed, the residual map from the blue channels shows positive residuals on the left side and negative residuals on the right side, while the opposite takes place in the residual map from the red channels. Therefore the line velocity is bluer than the average one on  the left side, and redder on the right side, as already suggested by the spectroastrometry. This is expected for a rotating disk with the line of nodes oriented E-W with the eastern half approaching the observer.

\begin{figure*}
 \centering
  \includegraphics[width=0.9\linewidth]{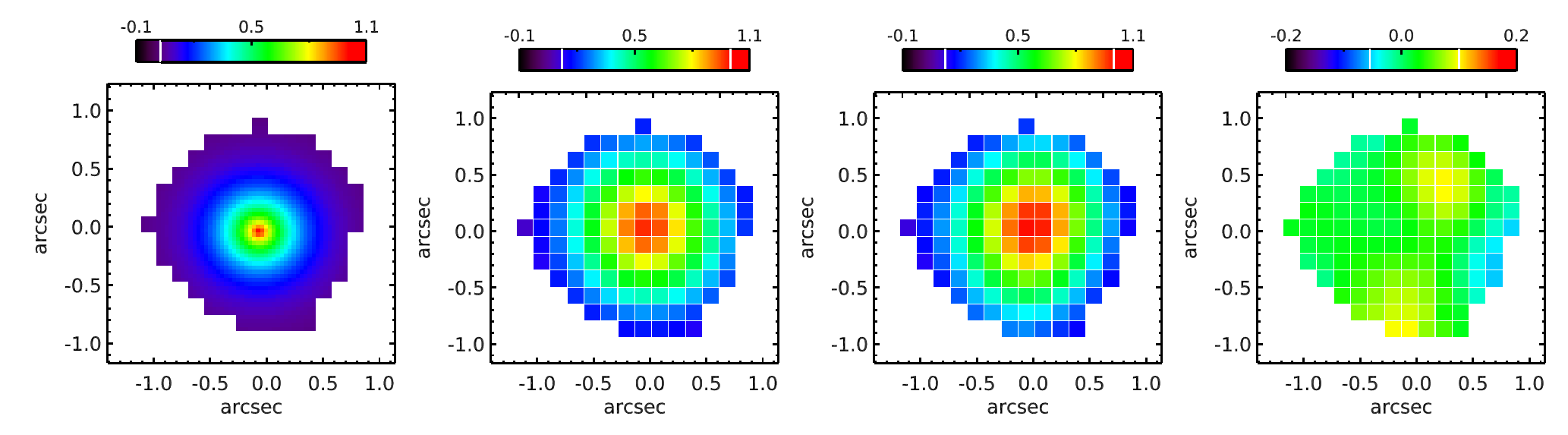} 
  \includegraphics[width=0.9\linewidth]{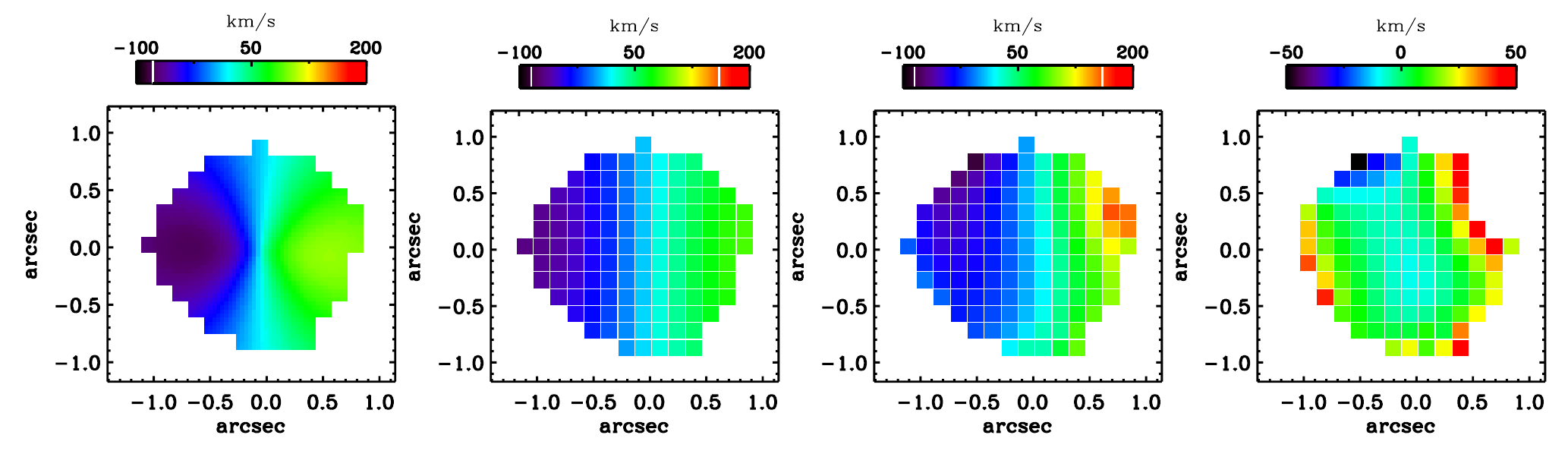} 
   \caption{Analysis of the SMG surface brightness and kinematics. Upper panels: surface brightness fitting with an exponential disk. Lower panels: velocity field fitting with a circularly rotating disk. From right to left: model, model convoluted for PSF, data and residual map.}
              \label{fig:SMG_model_fit}%
    \end{figure*}

The observed radial profile of the [CII] surface brightness and the two tests performed above strongly suggest that the [CII] kinematics has been spatially resolved. When extracting the line kinematics we took into account that the  [CII] line profile is the result of the combined emission of the SMG and of its faint companion observed in the PV diagram at a velocity of $\sim$250 km~s$^{-1}$. Since the emission of the companion is probably too weak to resolve its kinematics, we performed a pixel-by-pixel  fit with the two Gaussian components, but allowing only the stronger one to vary its average velocity and velocity dispersion. Therefore we  obtained the velocity field only for the SMG galaxy, cleaning out the distortions due to the faint component. The velocity and flux maps, shown in figure \ref{fig:SMG_model_fit}, were obtained after selecting only those spatial pixels for which the signal-to-noise ratio of the emission line is equal to or higher than 3; we defined the signal-to-noise ratio of the emission line as the ratio between the peak of the main Gaussian component and the rms of the residuals.


We then fit the velocity map of the SMG using the kinematical disk model used for $z\sim 3$ Lyman-break galaxies by  \cite{gnerucci:2011a}. We assumed that the gas is circularly rotating in a thin disk, that the gravitational potential depends only on  disk mass, and that the disk surface mass density distribution is exponential $\Sigma(r)=\Sigma_0 e^{-r/r_0}$, where $r$ is the distance from the disk center and $r_0$ is the scale radius. Following \cite{cresci:2009} and \cite{gnerucci:2011a}, we also assumed that the mass density distribution is traced by gas emission, and therefore that  the center of rotation and  scale radius are set by the peak and  scale radius of the [CII] surface brightness distribution parameterized by ${\rm I}(r)={\rm I}_0e^{-r/r_0}$ \citep{cresci:2009}.
 Rotation center and $r_0$ were then derived by $\chi^2$ minimization following the comparison of  observed and  model  surface brightness distributions after convoluting the latter with the spatial PSF of the data. 
The circular velocity field was  projected along the line of sight, weighted by the intrinsic line surface brightness, and then convolved with the  spatial resolution of the data to be directly compared with the observed velocity field. 

The free parameters of the kinematical model are $i$, the inclination of  the disk along the line of sight, $\theta$, the position angle of the disk line of nodes, $V_{sys}$, the systemic velocity of the galaxy, and $M_{\rm dyn}$, the total  dynamical mass of the disk. 
$M_{\rm dyn}$ is actually the disk mass enclosed within a radius of 5 kpc. Since this radius is much larger than the disk scale radius, this corresponds to the total mass of the disk.
The value of 5 kpc was chosen because the proper distance at $z\sim 5$  corresponds to a proper distance of 10 kpc at $z\sim 2$, which is the value used in the work of  \cite{cresci:2009}. 

 The best-fit parameter values for  the kinematical model were found using  the MPFIT \citep{markwardt:2009} and AMOEBA algorithms in the IDL environment. The $\chi^2$ space was  explored using  Monte Carlo Markov chains (MCMC), which allow one to sample a target density distribution, or, in other words, allow one to estimate the posterior probability distribution for the $i$-dimensional parameter space that defines our disk models. We used the IDL routines developed for the EXOFAST IDL package \citep{Eastman:2013}. 

Since the disk inclination is not well constrained from the data and since the dynamical mass could be larger than the molecular gas mass, we considered  two cases, one with  a totally free $M_{\rm dyn}$ and the other with a prior on $M_{\rm dyn}$ based on the molecular gas mass. The prior is given by a lognormal Gaussian step function, that is a lognormal Gaussian on the left side of the peak and a constant value on the right. The peak of the Gaussian is given by the molecular gas mass presented  in  table \ref{tab:properties}, and the dispersion is $0.3$ dex, corresponding to a factor 2 and compatible with the uncertainties on the conversion factor (e.g.~\citealt{bolatto:2013}).

\begin{figure}
 \centering
  \includegraphics[width=1.\linewidth]{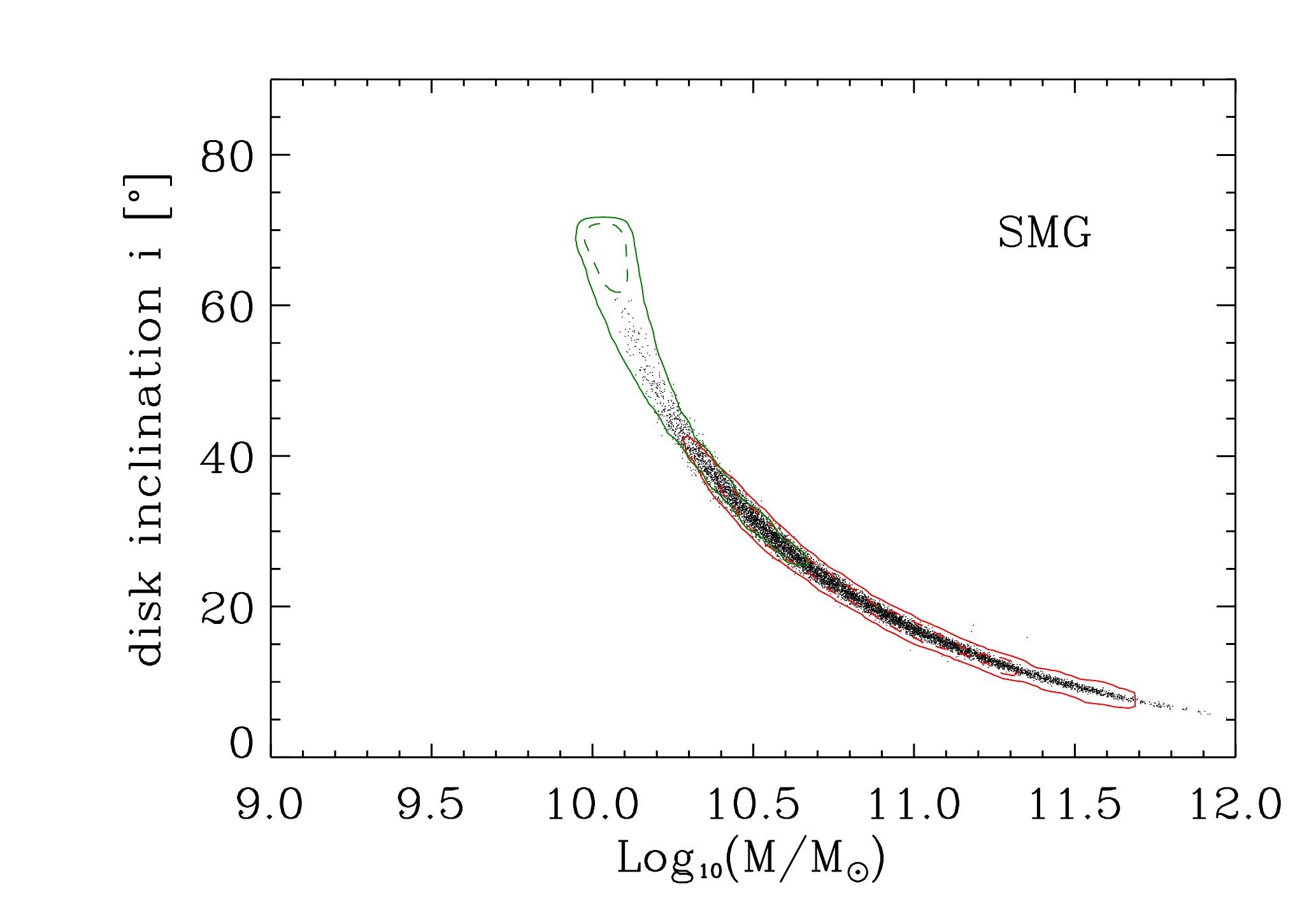} 
   \caption{Confidence contours for the dynamical mass and inclination values of the SMG obtained with the MCMC and 10$^5$ trials. The solid and dashed lines identify the 95\% and 60\% confidence limits, respectively. The red and green contours denote the cases with and without the prior on the dynamical mass, respectively.}
              \label{fig:SMG_Mdin}
    \end{figure}

Figure \ref{fig:SMG_model_fit} shows the results of the kinematical model fitting and table \ref{tab:properties} reports the best-fit parameters. Figure \ref{fig:SMG_Mdin}   show the confidence contours for the M$_{\rm dyn}$ and inclination $i $ values obtained from the MCMC  with  $10^5$ trials with (red) and without (green) the  dynamical mass prior. 

The scale radius here is $r_0 = 2.8\, \mathrm{kpc}$, similar to the maximum distance with measured velocity from the disk center,  indicating that the observed velocity field traces only the inner regions of the galaxy where dark matter is very likely  negligible. It also shows that deviations from the exponential mass density profile due to dark matter are very likely negligible and that $M_{\rm dyn}$ is indeed close to the total disk mass because $2 r_0 \sim 5$ kpc. 

It is clear from figure \ref{fig:SMG_Mdin} that without the prior the best fitting dynamical mass would be smaller than the gas  molecular mass by $\sim 0.5$ dex. However, the velocity map of the SMG is limited to the central parts of the disks and, given the spatial resolution, it lacks a clearly defined spider diagram: only the central velocity gradient is well detected. Therefore, the inclination cannot be well constrained from the data: given the well-known degeneracy of mass and inclination,  the mass is poorly constrained, too. The use of the prior therefore provides  a physically motivated 
lower limit to the dynamical mass, which helps constraining its value.

We  also performed a fit without the proxy on mass and in which the disk inclination was constrained by assuming that the [CII] flux distribution was radially symmetric on the disk: the resulting disk inclination is in the range $i=6^\circ\pm3^\circ$ (90\% confidence), which additionally supports the low disk inclination.

To verify whether the rotating disk is really rotationally supported, we compared the intrinsic velocity dispersion with the rotational velocity.
By intrinsic velocity dispersion we mean the one that is not accounted for by unresolved rotation and which might be due to turbulent motions.
From the dynamical modeling we can estimate the maximum rotational velocity (V$_{\rm max}$ ) and the intrinsic gas velocity dispersion $\sigma_{\rm int}$ 
$$
\langle\sigma_{\rm int}^2\rangle = \langle\sigma_{\rm obs}^2 -\sigma_{\rm model}^2\rangle
$$
where $\sigma_{\rm obs}$ is the observed velocity dispersion  and $\sigma_{\rm model}$  is that expected from the model that best fits the velocity map. $\sigma_{\rm obs}$ is directly obtained from the main Gaussian line component without any correction for the spectral resolution; $\sigma_{\rm model}$ is computed by assuming  a cold rotating disk, that is ~ with no intrinsic velocity dispersion, and by taking into account the spectral and spatial resolutions of the observations. In particular, the main contribution  to $\sigma_{\rm model}$ is  given by  beam smearing, while the effect of the spectral resolution is negligible since line widths are large. $\sigma_{\rm int}$ is therefore the intrinsic velocity dispersion of the emitting material.  

 The ratio ${\rm V_{max}}/\sigma_{\rm int}$ was used to indicate whether  the kinematic of the galaxy is rotation dominated (V$_{\rm max} > \sigma_{\rm int}$) or dispersion dominated  (V$_{\rm max} < \sigma_{\rm int}$). For the SMG we obtained ${\rm V_{max}}/\sigma_{\rm int}\sim 1.3$ with ${\rm V_{max}}\sim270$ km/s and $<\sigma_{\rm int}> \sim 200$ km/s. This indicates that the submillimeter galaxy has a dynamically hot disk compared with local galaxies for which 
  ${\rm V_{max}}/\sigma_{\rm int}\sim 10$, but comparable with LBGs at redshift $z\sim 2-3$  \citep{cresci:2009,gnerucci:2011a}. 

Summarizing, the mass and inclination of the SMG disk were derived from the posterior distributions of the parameters by considering the median values and the 5 and 95 percentiles and result  in $Log_{10}({\rm M}_{\rm SMG}/ {\rm M}_\odot) =10.8 \pm 0.6 $ (${\rm M}_{\rm SMG} =6^{+18}_{-5} \, \times 10^{10} \, {\rm M}_\odot)$  and $i = 25^\circ\pm 15^\circ$ .

The spectroastrometric estimate performed at the beginning of this section $M_{\rm SMG,vir}= (4.7\pm 1.5)\times10^{10}$ \ \msun \ (for an average inclination value of $i=60\deg$, taking into account the 0.15 $dex$ systematic error) is perfectly consistent with the dynamical mass measured from  the spatially resolved kinematics, confirming the reliability of the spectroastrometric method of \cite{gnerucci:2011}.

We conclude this section by roughly estimating the amount of mass that could be supported by the disk velocity dispersion, because ${\rm M}_{\rm SMG}$ is only the amount of mass that is supported against gravity  by disk rotation.
If the velocity dispersion indeed provides support against gravity, the dispersion-supported mass would not be negligible since $\sigma_{int}\sim V_{rot}$. In this case, unfortunately, it is not possible to apply simple recipes such as the asymmetric drift correction (see, e.g., \citealt{barth:2001}), and we  therefore rely only on a virial estimate. With $\sigma_{int}\simeq 200$ km/s and $r_0=2.8$ kpc the dispersion-supported mass is $M_{SMG,disp}\simeq \sigma_{int}^2 r_0/G = 2.6\times 10^{10}\msun$, similar to the rotation-supported mass, as expected.

\subsection{The quasar BR1202 South}

   \begin{figure*}
   \centering
\includegraphics[width=0.8\linewidth]{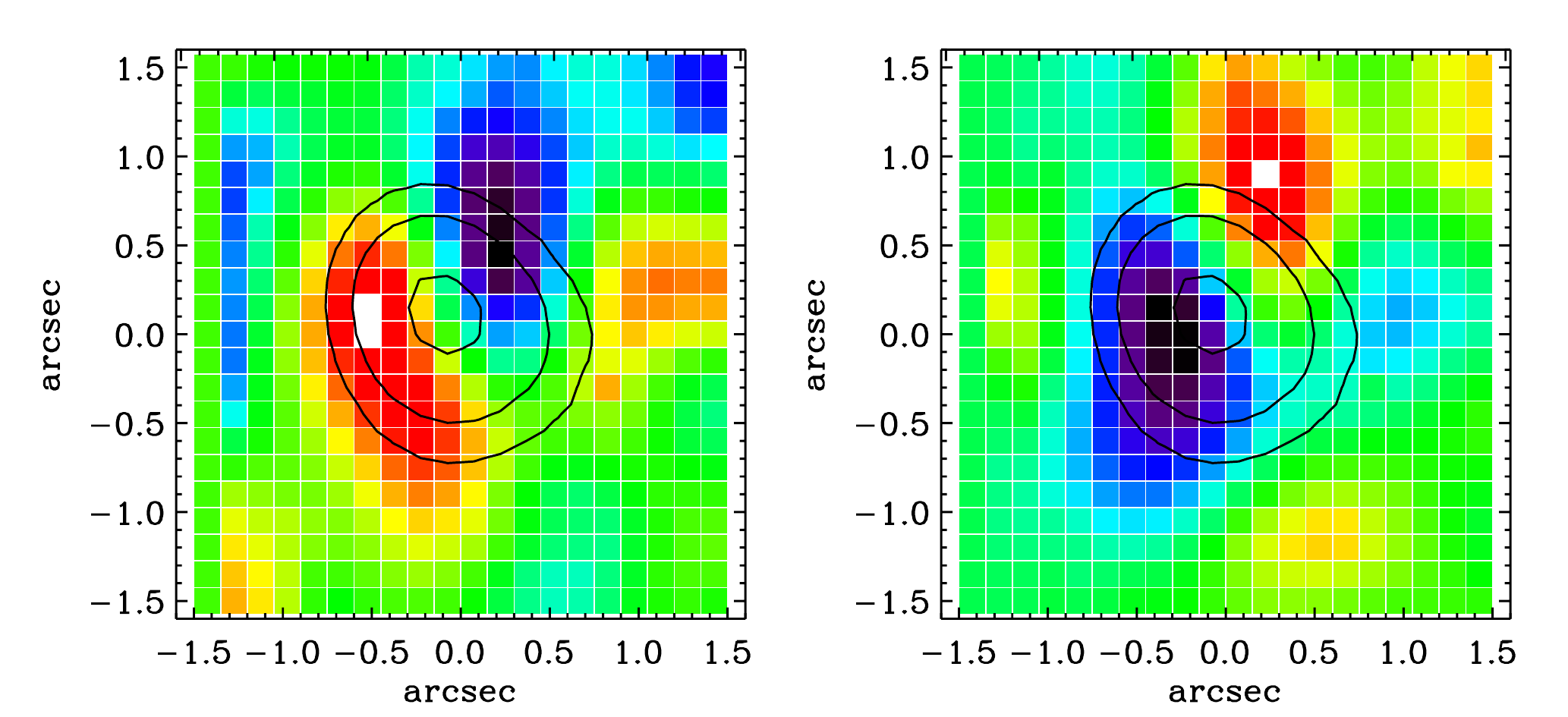}
      \caption{Blue and red residual maps of the QSO host galaxy obtained assuming spatially unresolved kinematics (left and right panels, respectively).  Contours   represent the total line surface brightness at 90\%, 50\%, and 30\% of the peak value. 
              }
         \label{fig:residual_map_QSO}
   \end{figure*}

 \begin{figure*}
 \centering
 \includegraphics[width=0.9\linewidth]{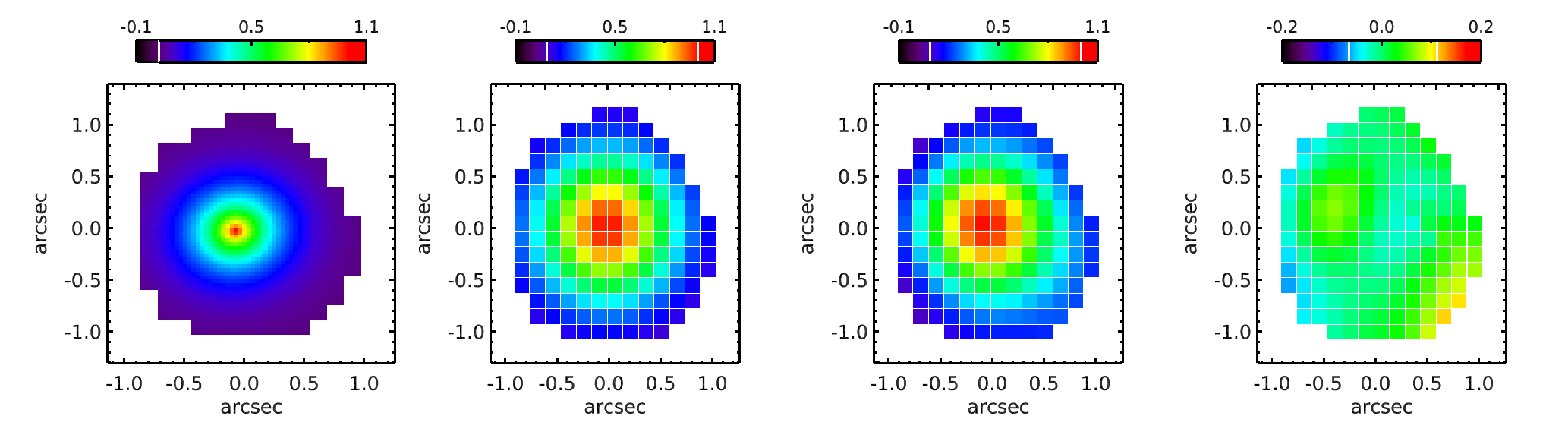} 
  \includegraphics[width=0.9\linewidth]{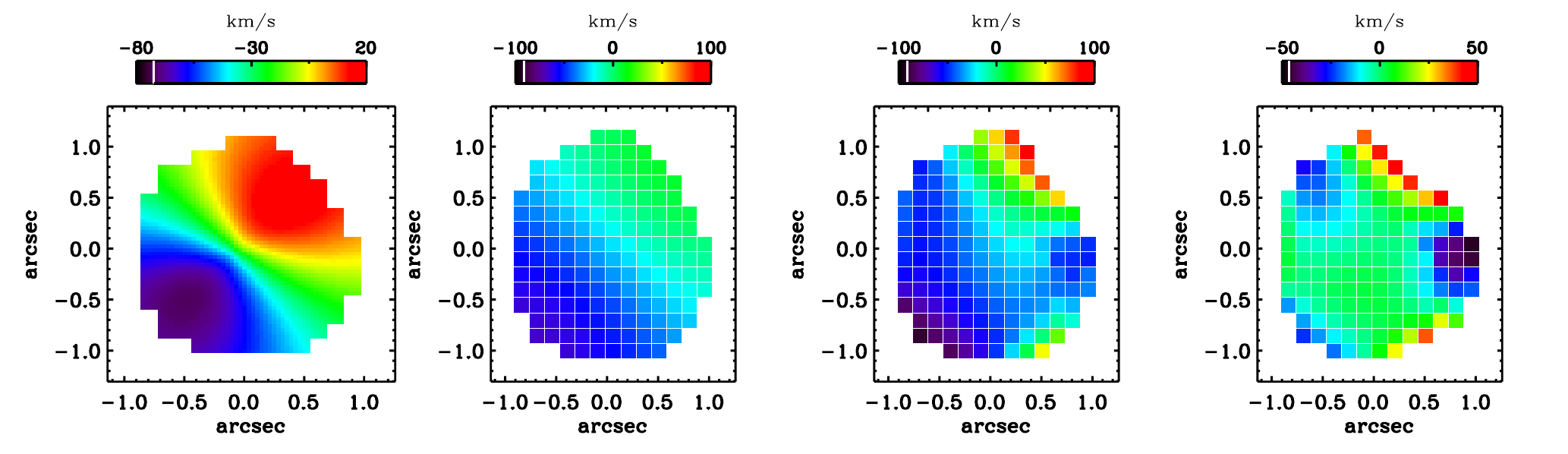} 
   \caption{Analysis of the QSO surface brightness and kinematics. Upper panels: surface brightness fitting with an exponential disk. Lower panels: velocity field fitting with a circularly rotating disk. From right to left: model, model convoluted for PSF, data and residual map. }
              \label{fig:QSO_model_fit}%
    \end{figure*}

The  [CII] position-velocity diagram  for the QSO (fig. \ref{fig:position_velocity}, right panel) 
was obtained by extracting a spectrum from a three-pixel-wide (0.45\arc) slit centered on the surface brightness peak and oriented N-S to compare with \cite{salome:2012}.
Similarly to what was found in the PV diagram of the CO(5-4) line by \cite{salome:2012}, the two main components at $v=-30$ and $40$ km/s a rotating gas disk. 

The results of the spectroastrometric test, performed in the same way as for the SMG, are shown in fig.~\ref{fig:spectroastrometry} and confirm the possibility of a rotating disk.
The dynamical mass estimated from spectroastrometry following \cite{gnerucci:2011a} is $M_{\rm dyn} \sin^2 i = (8.3\pm1.0)\times10^{9}$ \msun for measured values of $FWHM = 300\pm20$ km/s and $r = (0.4\pm0.1)$ kpc, the half distance between the blue and red surface brightness centroids.

The results of the residuals tests are shown in figure~\ref{fig:residual_map_QSO}  and, as for the SMG, they confirm that the kinematics is spatially resolved and that the disk is probably rotating.

The [CII] kinematical map for the quasar host galaxy has been obtained with a pixel-by-pixel fitting of two Gaussian functions, as suggested by the presence of the red wing shown in figure \ref{fig:spectrum}. As for the SMG galaxy, we first fit the spectrum obtained from the integrated QSO emission with  two Gaussian components, constrained to have the same line width. The fainter component, which reproduces  the red wing, was  then kept fixed in the pixel-by-pixel fitting except for its flux.
The resulting surface brightness and velocity maps are shown in  figure \ref{fig:QSO_model_fit}. As for the SMG galaxy, the QSO velocity field shows a velocity gradient oriented in the same direction as the shift between the red and blue centroids. 
To ensure that the kinematics of the QSO host galaxy is not affected by our decomposition method, we furthermore assumed that the red wing was caused by outflows  or tidal interactions
and  therefore is emitted by gas with a high velocity dispersion \citep{Carilli:2013}. After constraining the width of the fainter Gaussian function to  $\sim$ 400 km/s  \citep{Carilli:2013}, 
we repeated the previous procedure and obtained a kinematical map that is consistent with the former, within the errors.

\begin{figure}
 \centering
  \includegraphics[width = 1. \linewidth]{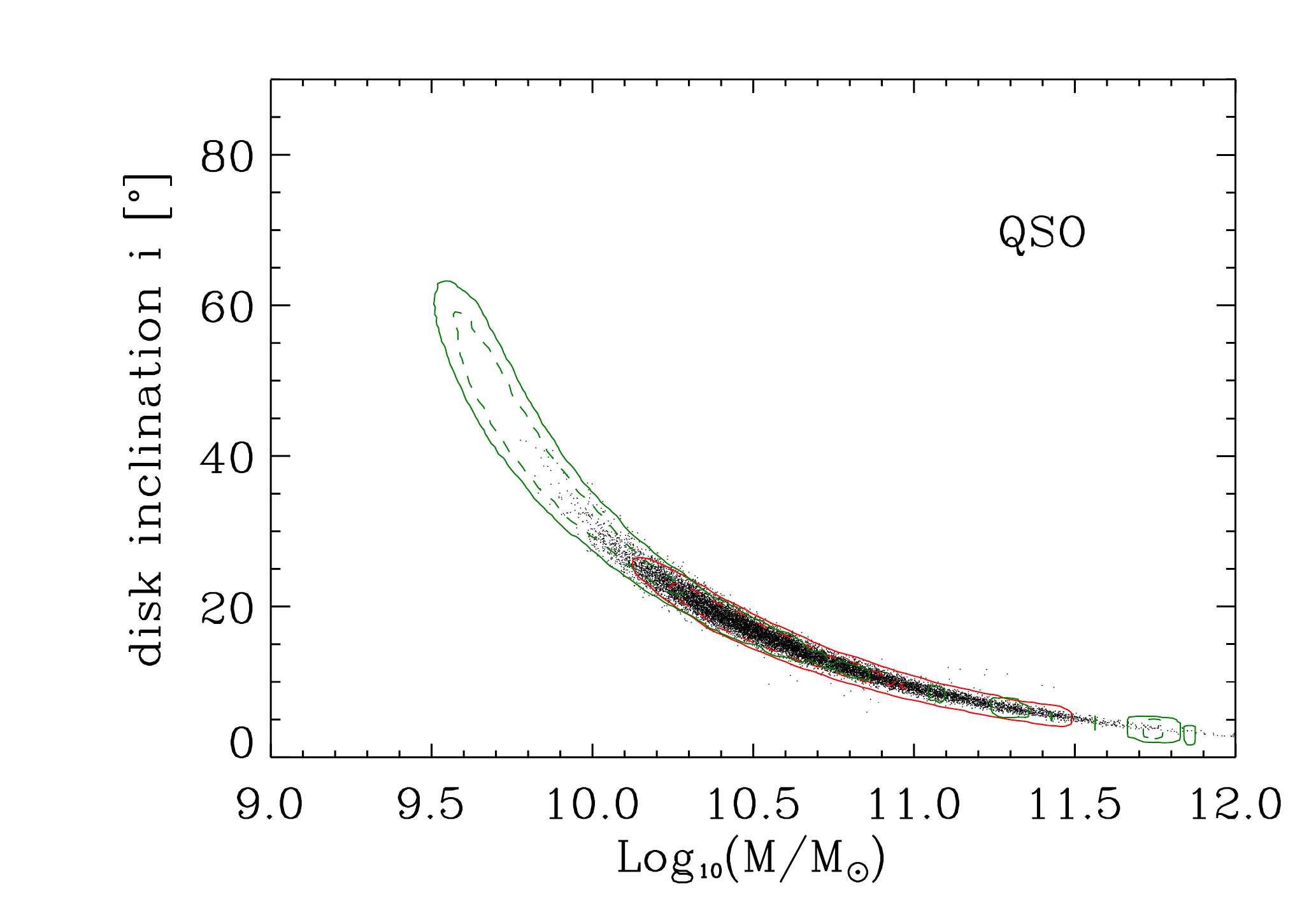} 
   \caption{Confidence contours for the dynamical mass and inclination values of the QSO obtained with the MCMC and 10$^5$ trials. The solid and dashed lines identify the 95\% and 60\% confidence limits, respectively. The red and green contours denote the cases with and without the prior on the dynamical mass, respectively.  }
              \label{fig:QSO_Mdin}
    \end{figure}

We then repeated the analysis performed for the SMG galaxy and fit the QSO kinematics with a rotating disk.  
We computed an MCMC with $10^5$ trials with and without the molecular gas mass prior; the results are shown in figure 
\ref{fig:QSO_Mdin}, where, as before, the red and green contours denote the cases with and without the prior on the mass. At variance with the case of the SMG, the confidence contours on the two cases overlap at $\log{M/M_\odot}\sim 10.6$, indicating that even without the prior, the dynamical mass is consistent with the molecular gas mass.
All the relevant best-fit parameter values are also shown in table \ref{tab:properties}.

In this case we also   performed a fit without the proxy on mass and with the disk inclination  constrained by assuming that the [CII] flux distribution is radially symmetric on the disk: the resulting disk inclination is in the range $i=2^\circ\pm2^\circ$ (90\% confidence), which also supports the low disk inclination, as for the SMG.

As in the case of the SMG, by using the rotating disk model we estimated ${\rm V_{max}}\sim 230$  km/s and a $\langle{\rm \sigma_{int}}>\rangle\sim 150$ km/s and corresponding ratio ${\rm V_{max}}/{\rm \sigma_{int}}\sim1.5$. This indicates a dynamically hot disk compared   to local galaxies, but still marginally colder than that of the SMG.

Summarizing, the mass and inclination of the QSO disk  derived from the posterior distributions of the parameters by considering the median values and the 5 and 95 percentiles result in $Log_{10}({\rm M}_{\rm QSO}/ {\rm M}_\odot) =10.6^{+0.8}_{-0.4} $ (${\rm M}_{\rm QSO} = 4^{+20}_{-3} \, \times 10^{10} \,{\rm M}_\odot)$)  and $i = 15^\circ\pm 10^\circ$. 
 This is consistent within the error with the mass estimated from spectroastrometry.

Finally, using $\sigma_{int}\simeq 150$ and $r_0 = 2.4$ kpc, the dispersion-supported mass is $M_{QSO,disp}\simeq \sigma_{int}^2 r_0/G = 1.2\times 10^{10}\msun$, a significant contribution to the rotation-supported mass, as expected.

Overall, the total SMG and QSO mass is smaller than the lower limit we estimated for the dark halo mass.

Intriguingly,  the QSO and SMG have similar SFRs (between 1000 and 5000 \msun\ for both), molecular gas masses ($\sim 4\times 10^{10}\,$\msun) and moderately similar dynamical masses when the different inclinations of the gas disks are considered. 

   \begin{figure*}
   \centering
\includegraphics[width=0.4\linewidth]{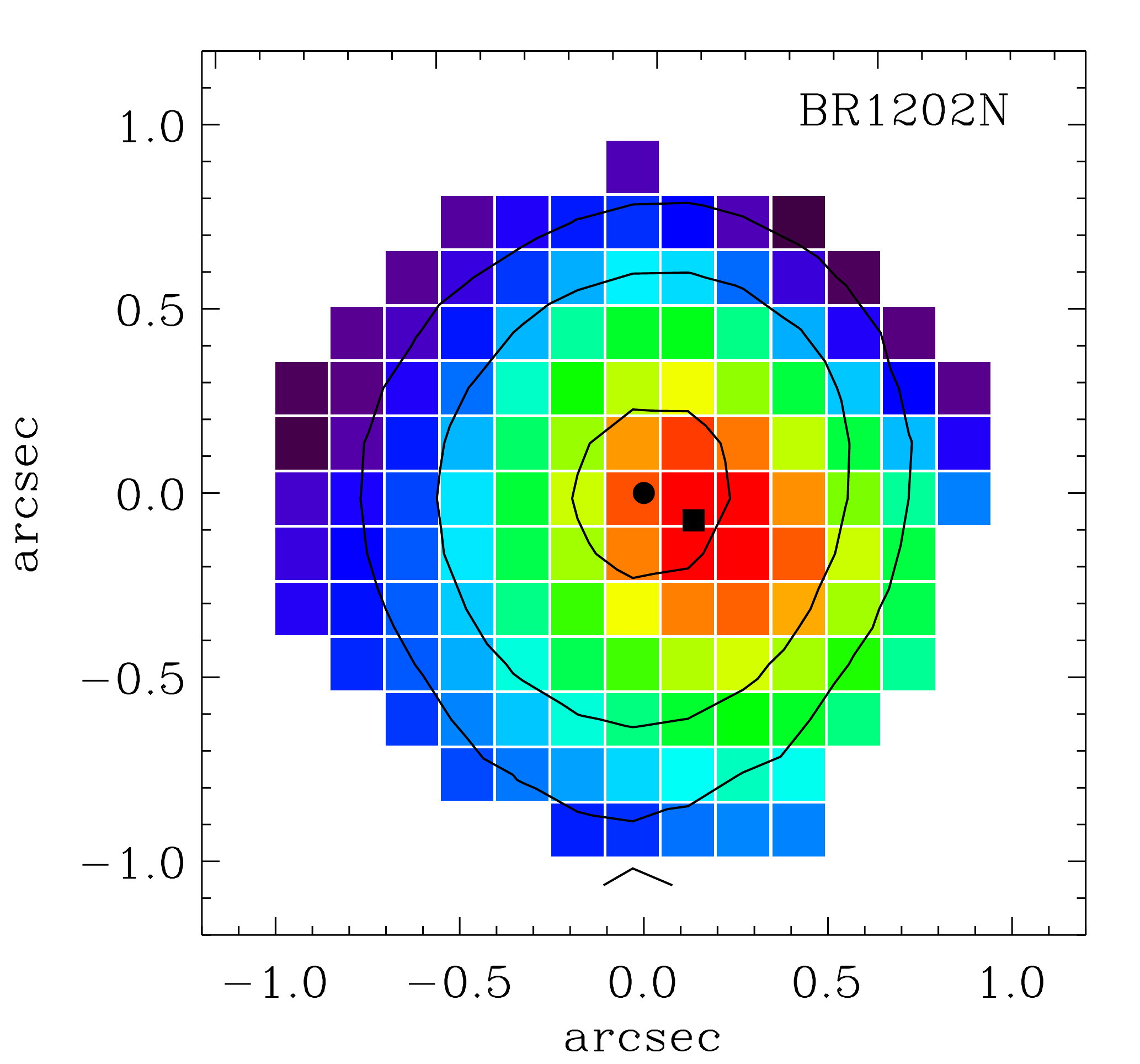}
\includegraphics[width=0.4\linewidth]{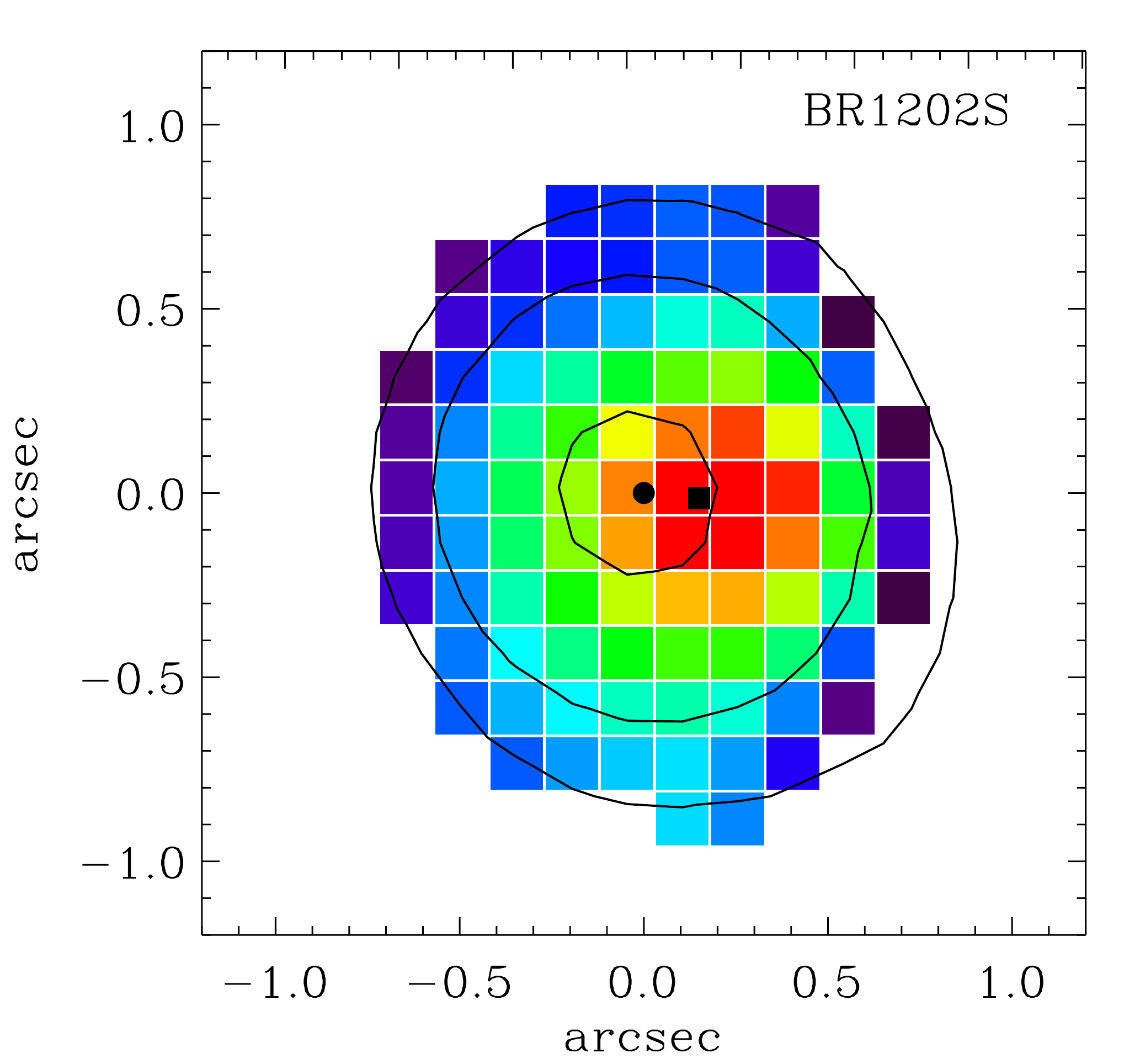}
      \caption{ Surface brightness maps of the sources  identified as faint components in the spectral fits:  the left and right panels show the SMG and QSO, respectively. The black contours denote the surface brightnesses of the main spectral components; the filled circles and squares denote the centroids of the main and faint components, respectively.   }
         \label{fig:faint_comp}
   \end{figure*}

   \begin{figure*}
   \centering
\includegraphics[width=0.8\linewidth]{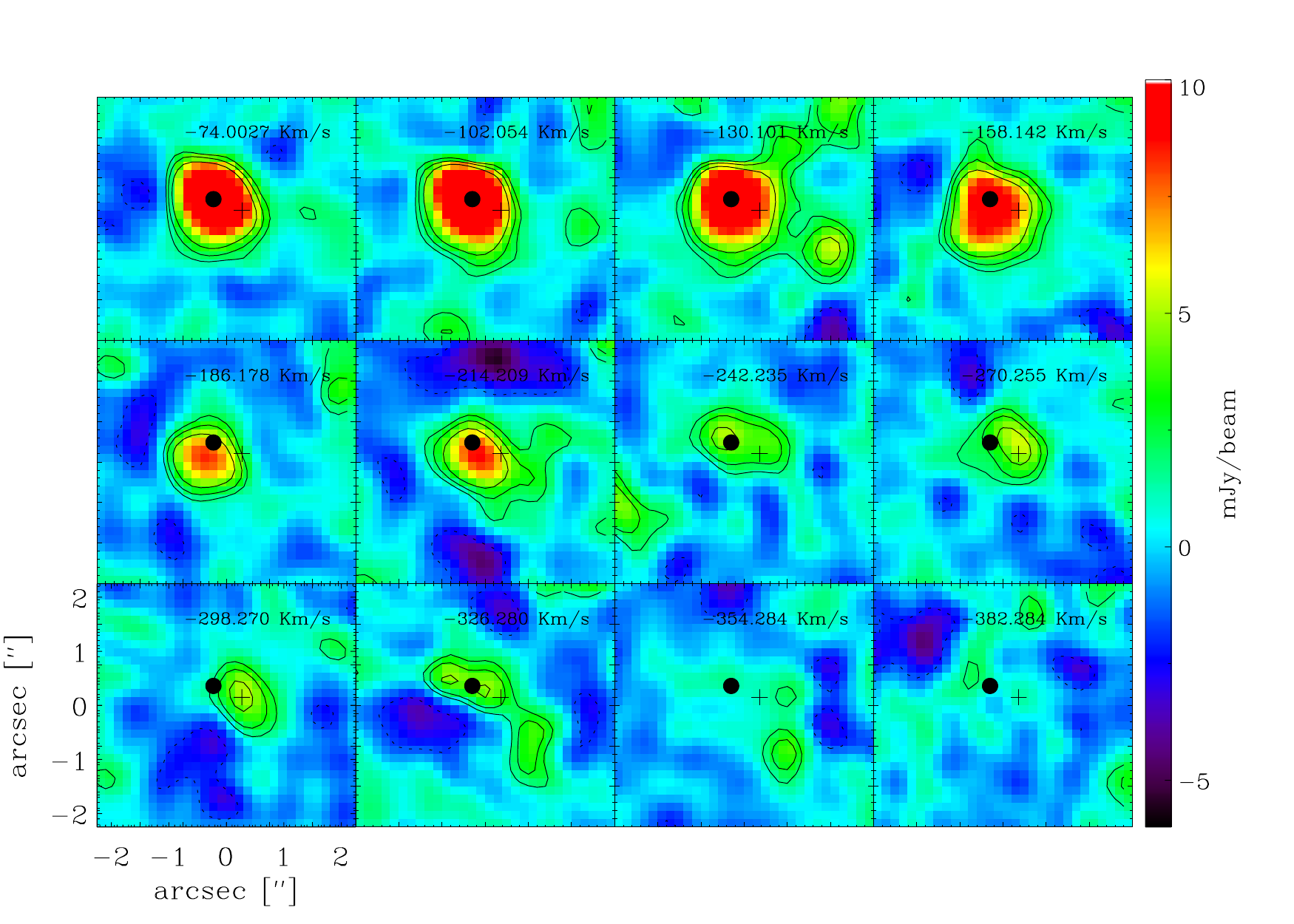}
      \caption{ Brightness per velocity channel in the QSO. Color-coded values and contours at [2,3,5] $\sigma$ levels are shown. The numbers in the upper right corners of the panels denote the central velocity of the channel. The filled circle represents the peak of the integrated surface brightness of the quasar, while the cross  represents the peak of  the surface brightness of the faint companion, which is detected in channels $\sim-362:-242$ km/s.  }
         \label{fig:channel_map}
   \end{figure*}

\subsection{Spectroscopically detected faint companions} 
 In figure \ref{fig:faint_comp} we show the surface brightness maps of the sources identified as faint Gaussian components in the spectra plotted in figure \ref{fig:spectrum}. These components have velocities of 570 km/s (SMG) and 500 km/s (QSO) with respect to the main components, and velocity dispersions of 220 km/s and 150 km/s. These values   where kept fixed during the  pixel-by-pixel fit. The surface brightness centroids of these components have a distance of 0.24\arcsec\ and 0.15\arcsec\  from the SMG and QSO, respectively, which at the redshift of the BR1202-0725 system corresponds to $\sim1.6$ and $\sim0.9$ kpc. These components are only detected spectroscopically, and because of their faintness and  distance from the main components cannot be seen in  direct images obtained by collapsing a given velocity range in the datacube. Therefore these are sources additional  to those identified by \citealt{wagg:2012} and \citealt{Carilli:2013}. 

 Figure \ref{fig:channel_map} shows the [CII]  channel map for the QSO where, in the channels at velocities from -326 to -242 km/s, we observe  a small weak component whose integrated surface brightness peak is indicated by a  cross. For comparison, the filled circle indicates the centroid of the QSO main emission.
 Its relative velocity with respect to the quasar is $\sim -250$ km/s for a projected distance of $r=3.4$ kpc. 
 

Overall, we have  spectroscopically identified three new components at distances less than 3.5 kpc from the SMG and QSO host galaxy. Considering also the two components identified by \citet{wagg:2012} and \citet{Carilli:2013} which correspond to the Ly-$\alpha$ emitters  visible in the HST images, we have a total of at least five faint components on top of the two main components, which are the SMG and QSO host galaxy. The BR1202 system  clearly represents a complex system, probably a dense group or protocluster in the early Universe with intense ongoing star formation. 
However, it is not possible to assess whether these components represent small galaxies, or gas clouds in tidal tails, inflows or outflows. They are probably they are the same type of objects as Ly-$\alpha$ emitters. 

However, assuming that they are subjected only to gravitational motions, it is possible to use these components for a rough estimate of the mass of the associated main component (SMG or QSO). Assuming that their orbital planes are seen edge-on and that their motions are entirely along the line of sight, we can combine their velocities and distances from the main components to obtain a virial mass estimate of the SMG or QSO host galaxy.  Given our assumptions, these are clearly lower limits on the true mass values. The inferred masses are $\sim 1.2\times10^{11}$ \msun \ for the SMG and $\sim 5.2\times10^{10}$ \msun , $\sim 5.9\times10^{10}$ \msun \ from the redshfited and blueshifted components of the QSO. These lower limits should be compared with the measurements performed in the previous sections, which are $0.6^{+1.9}_{-0.4}\times10^{11}$ \msun\ for the SMG and $0.4^{+2.1}_{-0.2}\times10^{11}$ \msun\  for  the QSO. These values are evidently agree  in remarkably  with the measurements from the spatially resolved kinematics supporting the hypothesis of gravitational motions for the spectroscopically detected components.
However, it should be noted that (i) any correction for the unknown geometrical projection effects will increase the mass estimate, (ii) the SMG and QSO host galaxy masses enclosed within the measured distances from the faint components are smaller than the total dynamical estimates, and (iii) the above estimates are based on the assumption of circular orbits while, higher velocities could be found with other, more general, types of orbits.
Both (i) and (ii)  will undoubtedly worsen the agreement with the dynamical masses at an unknown level, while (iii) would improve it. 

It is also possible that these faint components represent quasar- or star formation-driven outflows. This interpretation has been proposed by  \cite{Carilli:2013} for the red wing of the QSO.
However,  outflows detected from emission lines at submm wavelengths are seen on both blue and red sides of the emission line profiles (e.g., \citealt{feruglio:2010,maiolino:2012a}). Moreover,  it would be very peculiar that we are seeing almost face-on galactic disks for the SMG and QSO  while detecting only outflows on the far sides.

Summarizing, although it is not possible to assess the physical origin of the faint spectral  components,  the relative velocities and positions of the faint components detected spectroscopically suggest that they are orbiting in the gravitational wells of the SMG and QSO host galaxy, favoring their identification as small galaxies, or gas clouds in an accretion flow or in tidal stream. The relative velocities and positions  provide no support either for their identification as gas clouds in QSO or star formation driven outflows.
ALMA observations with higher sensitivity and spatial resolution are needed to establish the physical origin of these faint spectroscopic components.

\section{Black hole mass of the QSO}
   \begin{figure*}
   \centering
\includegraphics[width=0.8\linewidth]{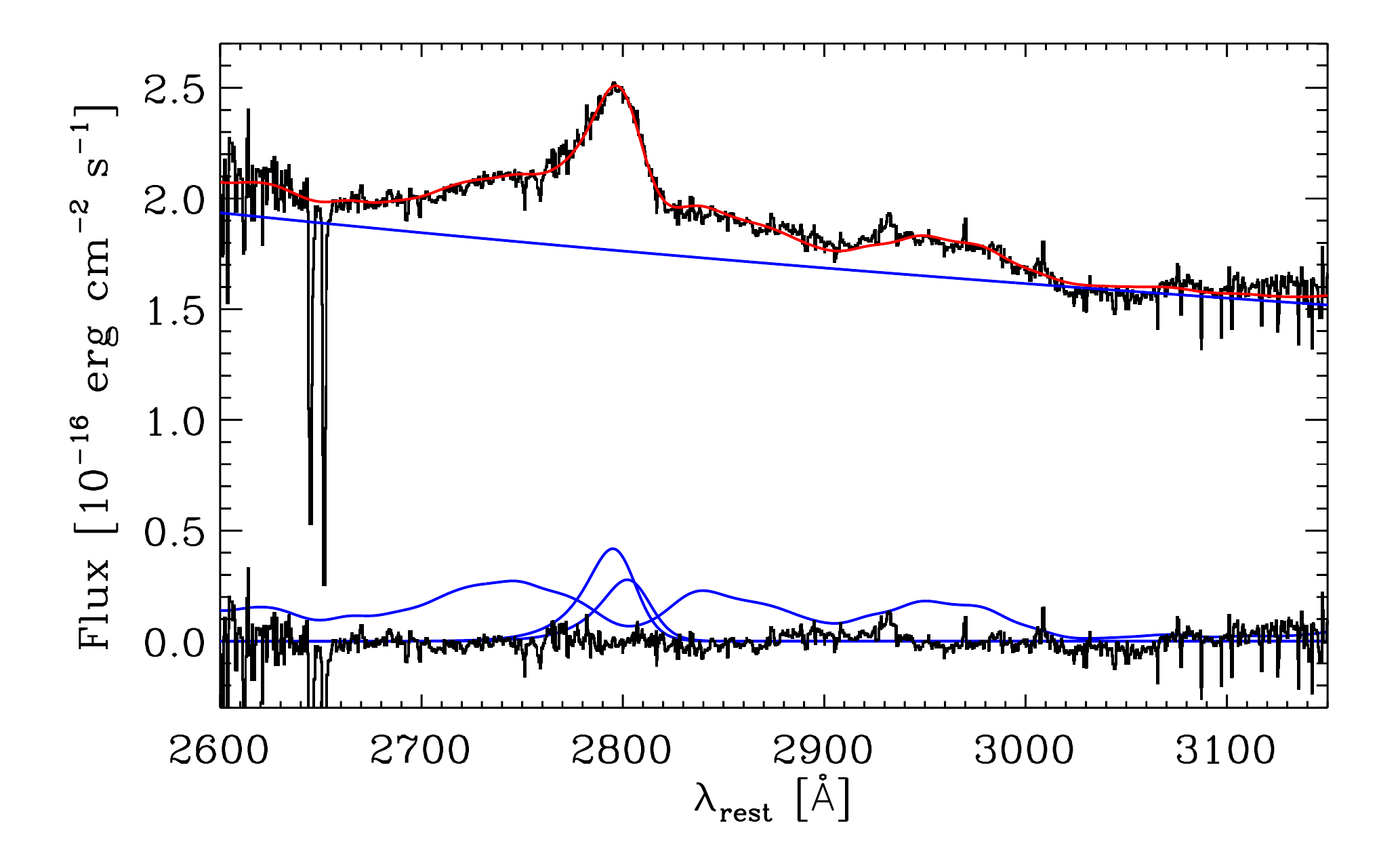}
      \caption{X-Shooter spectrum of the QSO (S source) and model fit (red line) in the good atmospheric regions.
      The MgII doublet, FeII spectrum and power law continuum are shown with blue lines. Below is the  residual spectrum.}
         \label{fig:mgii_fit}
   \end{figure*}

To estimate the  black hole mass in the QSO and compare it with the mass of the host galaxy, the only possibility is to use a virial estimate based on broad emission lines such as H$\beta$, MgII $\lambda 2800 $ \AA, \ and CIV $\lambda 1546$ \AA.  H$\beta$ and  MgII are currently  believed  to be  the most accurate BH mass estimators with systematic uncertainties of up to $\sim 0.5 $ dex, while CIV is very likely affected by outflows or nonvirial motions (see \citealt{Shen:2013} for a review). 

An intermediate-resolution spectrum of the QSO has been obtained with X-shooter \citep{vernet:2011} at the ESO VLT 
(P.ID. 084.A-0780B) with the main aim of studying galactic and intergalactic absorption lines.
A total of ~2.5 hours of exposure time was obtained using a slit set of 0.8\arcsec/0.7\arcsec/0.6\arcsec\ for the blue, visual (VIS), and near-infrared (NIR) arms, resulting in spectral resolutions of 6000, 11000, and 7950, respectively.
The data were reduced with the public X-shooter pipeline \citep{modigliani:2010} using the default parameters. The pipeline rebins the spectra to 0.4 \AA/pixel in the blue and VIS arm and to 1 \AA/pixel in the NIR arm.
Telluric absorption was removed using the spectrum of  the B5V star Hip 073345 after removing the prominent H absorption lines.
The NIR portion of the spectrum (extending from 1 to 2.5 mum) includes the broad MgII emission line doublet at ~2800 \AA\ rest frame, which is considered a good estimator for  the mass of the supermassive black hole. The region of the spectrum around the MgII emission is shown in figure \ref{fig:mgii_fit}.

The spectrum was fit with a combination of power-law continuum, FeII emission and broad line. For the FeII emission we used a combination of model templates generated with CLOUDY (version 10.00; \citealt{ferland:2013}) at  the conditions typical of the BLR and allowing for Doppler broadening with a Gaussian of given velocity dispersion. For each of the MgII lines we used  a broken power-law profile \citep{nagao:2006} convolved with a Gaussian with unknown velocity dispersion. The two MgII lines were forced to have the same line profile in velocity, with a flux ratio $\lambda$2796/$\lambda$2803 limited in the 1.0-2.0 range \citep{laor:1997}. The red line in figure \ref{fig:mgii_fit}  shows the best model fit, which results in the following line width and continuum luminosity for the MgII lines: $FWHM =  3400\pm 100$ km/s and
$\lambda L_\lambda(3000\mathrm{\AA})= (1.1\pm 0.1)\times 10^{47}$ erg/s.
Extrapolating the power-law continuum, we obtain 
 $\lambda L_\lambda(1450\mathrm{\AA})= 1.3\times 10^{47}$ erg/s and $\lambda L_\lambda(5100\mathrm{\AA})= 9.5\times 10^{46}$ erg/s. 
 
The black hole mass was estimated using the virial relation $M_{BH} = f V^2 R / G$, where $V$ is the line width (FWHM or $\sigma$) and $R$ is the broad line region radius derived from the well-known radius luminosity relation \citep{kaspi:2000,bentz:2008}.  The scaling factor $f$ was calibrated by imposing that virial masses agree with the relations between BH mass and host galaxy properties observed in the local Universe (see, e.g., \citealt{peterson:2011}). Here we adopted the very recent calibration by Marconi et al.\ (2013, in preparation), obtained by matching the virial BH masses of the AGN with  reverberation-mapping data  with the BH masses estimated from the velocity dispersions and luminosities of the  host galaxy spheroids, using the $M_{BH}-\sigma$ and $M_{BH}-L$ relations by \cite{mcconnell:2013}:   the adopted virial relation is then 
$M_{BH}/M_\odot = 10^{6.7} FWHM_{1000}^2 [\lambda L_\lambda(3000\mathrm{\AA})]_{44}^{0.5}$ where the subscripts 1000 and 44 indicate units of 1000 km/s and $10^{44}$ erg/s, respectively. 
The zero point of this relations has a $\sim 0.3$ dex systematic uncertainty and
 is lower by $0.15$ dex with respect to the zero point of the calibration by \cite{vestergaard:2009}.
Finally, the estimated black hole mass is M$_{\rm  BH} = 1.9\times10^{9}$ \ \msun\ ($\pm 0.3$ dex, systematic).
The quasar bolometric luminosity was estimated from the UV and optical luminosities.
Using the relation by \cite{mor:2012}, we derived $L_{\rm AGN} \sim 3.8\times 10^{47}{\rm erg/s}$ from  $\lambda L_\lambda(1450\mathrm{\AA})$, while we derived  
$L_{\rm AGN} \sim 5.7\times 10^{47}{\rm erg/s}$ adopting $L_{\rm AGN} \sim 6 \lambda L_\lambda(5100\mathrm{\AA})$ from \cite{marconi:2004}.
The Eddington ratio is therefore $L/L_{Edd}\sim 1.5$.
Finally, combining the BH with the dynamical mass for the host galaxy of $\sim4\times 10^{10}\, M_\odot$, the ratio between the black hole and dynamical galaxy mass is $\sim 0.05$; if we also consider the contribution from the velocity dispersion to the host galaxy mass, this ration drops to 0.04.
These values are  consistent with the upper envelope of the local $M_{BH}-M_{gal}$ relation (see, e.g,  figure 3 of \cite{sani:2011} and fig.~\ref{fig:mbh_mbulge} below).

\section{Discussion}\label{sec:discussion}

According to the  paradigm of  co-evolution between BHs and their host galaxies, large bursts of star formation and AGN activity should be induced by mergers or interactions, at least at large galaxy masses: gravitational torques cause gas inflows that fuel star formation and accrete onto supermassive black holes in galaxy nuclei. 
In this framework it is predicted that an SMG and a QSO  represent two distinct phases (see, e.g., \citealt{alexander:2012} for a recent review). The SMG  represents the early phases after a major merging of gas-rich galaxies when  star formation and BH accretion proceed at very high rates in a dust-obscured environment.  
When the active galactic nucleus is powerful enough, it can expel the gas from the host galaxy, quenching both star formation and black hole accretion, 
and we are left with a QSO shining in an otherwise passive galaxy 
(e.g.~\citealt{di-matteo:2005,hopkins:2010,hopkins:2006}). 

The  BR 1202-0725 system, composed of a submm galaxy (SMG, N Source) and of a quasar (QSO, S Source) at $z\sim 4.7$, is an ideal test bench for this paradigm. The physical properties of the submm galaxy  and of the quasar are summarized in table \ref{tab:properties}.
Overall, the two galaxies have remarkably similar FIR continuum, [CII] and CO line luminosities, which can be translated into similar SFRs and molecular gas masses. Moreover, they also have similar dynamical masses.

These two galaxies are highly gas rich ($M_{mol}\sim 4\times 10^{10}\,\msun$) and strongly star forming  
($SFR\sim 3000\,\msun/yr$) with dynamical masses within the inner 5 kpc of about  $M_{dyn}\sim 5\times 10^{10}\,\msun$ which suggests high gas fractions.
Taking into account the uncertainties on the disk inclinations and on the conversion factor,  the corresponding gas fractions are   constrained in the ranges 0.1-1.0 (0.7 average) for the QSO and 0.2-1 (0.8 average) for the SMG.
The average values in parenthesis are for the average disk inclinations of  i=25$^{\circ}$ for the SMG and i=15$^{\circ}$ for the QSO. Although weakly constrained, the data  suggest moderately high to high gas fraction values.

 The relative proximity of the SMG and QSO (projected distance of $\sim 24$ kpc) suggests that the two galaxies might  be in the early phases of a major merger that is taking place within a relatively massive halo: the relative velocities of the two galaxies combined with the projected distance allow one to obtain a lower limit to the mass of the host halo, which is $> 3.2\times 10^{11}\msun$, compared with a total mass of $ 1.0^{+3.8}_{-0.8}\times 10^{11}\msun$ for the QSO and SMG.
 
Summarizing, the SMG and QSO have very similar characteristics, and  the only  real difference is that the S source is a QSO, although  it is not possible to exclude the presence of a luminous obscured AGN even in the SMG. 
The intriguing results are that these   high SFRs are apparently not triggered by  major mergers and that there are no evident signs of feedback in action.

\subsection{Rotating disks and the origin of the high star formation rates}
At variance with the major-merger scenario, the kinematical analysis we have performed indicates that both galaxies have velocity fields that  are consistent with regular rotation, without indications of significant distortions like those expected from  major mergers, strong interactions, or AGN feedback.  For the merger scenario, this fact confirms that the two galaxies are still far enough away from each other not to be affected by significant tidal forces.

Both galaxies are experiencing star formation at rates that will exhaust the gas on very short timescales. 
The observed molecular gas masses and SFRs imply gas depletion time scales of $\tau_{\rm dep}={\rm M_{gas}/SFR} \sim 10^7$ yr,  similar with those of ULIRGs (e.g \citealt{tacconi:2006,bouche:2007}). 
If the SMG and QSO are in an early merger phase, their gas might be entirely converted into stars before they actually do merge. In fact, the relative velocity between the two sources is about 240 km/s which, combined with a projected distance of  23.5 kpc, suggests a merger timescale of $\sim10^8$ yr, longer than the gas depletion timescale. Overall, there are no signs of  major-merger induced star formation, and the stars in the galaxies might be already formed when the two galaxies are close enough to merge.

The question that naturally arises is what triggered the burst of star formation in the SMG and the QSO since, apparently, it is not  a major merger.

Both the SMG and QSO in the BR 1202-0725 system are experiencing high star formation rates (SFR$>$1000 \sfr \ ) in rotating disks, apparently excluding major mergers. However, both galaxies show fainter companions that are revealed spectroscopically and whose emission lines have been deblended  when measuring the kinematics (see also \citealt{salome:2012}). The velocities and positions of these companions relative to the SMG and the QSO are consistent with gravitational motions within the gravitational potential wells of the two largest sources. Furthermore, \cite{Carilli:2013} and \cite{wagg:2012} find evidence for the emission of a faint  southwest source which is probably interacting with the S source, and also an additional companion between the two galaxies. Therefore the BR 1202-0725 system might be a massive proto-cluster  where the high star formation activity is probably triggered by minor mergers or interactions that destabilize the gaseous disks, but that do not significantly affect the kinematics of the SMG and QSO galaxies.  
Alternatively, the high star formation rate might be triggered by  cold gas inflows from the halo, which, again, destabilize the gas disks gravitationally.
Intriguingly the trigger of star formation from minor mergers, interactions or disk instabilities might be consistent with the dynamically hot disks,  indicated by the observed ratios ${\rm V_{max}}/{\rm \sigma_{int}}\sim 1.3$ and 1.5 for the SMG and for the QSO respectively.
Moreover, both velocity maps show residuals at the positions of the possible faint companions after  subtracting  the rotating disk models. However, the sensitivity and spatial resolution of the data do not allow us to quantify this result.

Evidence that  processes such as accretion of pristine gas might induce star formation is starting to accumulate. For instance, \cite{cresci:2010} found evidence  that gas accretion might be the origin of metallicity gradients in high-redshift star-forming galaxies. In a recent study, \cite{Kaviraj:2013}  estimated that only $\sim$27 percent of the total star formation budget in a sample of 80 massive galaxies a z$\simeq$2 is due to major mergers. This suggests that other contributions to star formation activity such as cold flows and minor merger might be operating. 
The dynamically hot rotating disks are also found in 
LBGs at  $z \sim 1$-$2$ (e.g. Genzel et al. 2006; F\"orster Schreiber et al. 2006) and at $z\sim 3$ (Gnerucci et al. 2011), associated with the very high gas fractions that can destabilize the disk gravitationally (Tacconi et al. 2010; Daddi et al. 2010a).
 Future observations with ALMA at higher spatial resolution and sensitivity might help to reveal the trigger of the strong star formation activity in BR1202 and in other systems.

To explore the origin of the star formation activity in more detail, we located the two galaxies on the Schmidt-Kennicutt relation at high redshift.
With the disk scale radii estimated during the kinematic analysis, we estimated the SFR and gas surface densities, $\Sigma_{\rm SFR}$ and $\Sigma_{\rm gas}$ in the two sources. The values obtained for the SMG ($\Sigma_{\rm SFR}$ = 80 \msun/yr kpc$^{-2}$, $\Sigma_{\rm gas}$ = 1900 \msun  \ pc$^{-2}$) and QSO ($\Sigma_{\rm SFR}$ = 180 \msun/yr kpc$^{-2}$, $\Sigma_{\rm gas}$ = 1800 \msun \ pc$^{-2}$)  are perfectly consistent with the relation found by \cite{daddi:2010} for  starburst galaxies (SMGs and ULIRGs, see their figure 2), which form a different relation from that of local star-forming galaxies.

Summarizing, the lack of  major-merger-induced features combined with the high SFR per  unit mass suggest that star formation in the BR 1202-0725 system is probably triggered by interactions, minor mergers, or inflow of gas from the halo that increase the concentration of the gas and dust in the galaxy and destabilize the disks gravitationally.

\subsection{Relation between the accreting BH and its host galaxy}

First of all, we compared the AGN and star formation luminosities in the QSO with the active sources at similar redshifts observed by \cite{ mor:2012}. Using  the $\lambda L_\lambda(1450\AA)$ value estimated from the XShooter spectrum and combining it with the bolometric correction by \cite{ mor:2012}, we derive $L_{\rm AGN} \sim 3.8\times 10^{47}{\rm erg/s}$. When combining this with $L_{\rm SF}\sim 7\times 10^{46}{\rm erg/s}$, the QSO is located above the $L_{\rm SF} \sim L_{\rm AGN}^{0.7}$ relation observed for lower luminosity sources \citep{netzer:2009}, but  close to the  five z$\sim 4.8$ sources  observed by \citep{ mor:2012}: the location above the local $L_{\rm SF} - L_{\rm AGN}$ relation might suggests that the feedback mechanism is still not effective enough to reduce star formation activity (see the discussion in \citealt{netzer:2009,mor:2012}). Indeed, the regular kinematics observed for the QSO host galaxy suggest that the feedback process might not yet be affecting star formation.

For the SMG, we did not find any evidence for an accreting BH,  but the presence of an obscured AGN is supported by radio and X-ray observations. The radio continuum flux density of the N source at 1.4 GHz  suggests the presence of a highly obscured AGN  \citep{carilli:2002,iono:2006}. This presence is tentatively  confirmed by the X-ray emission observed by Chandra at the SMG position \citep{iono:2006}. However, nothing more can be inferred, apart that from here we did not find any evidence for feedback, either.

\begin{figure}
 \centering
  \includegraphics[width=1\linewidth]{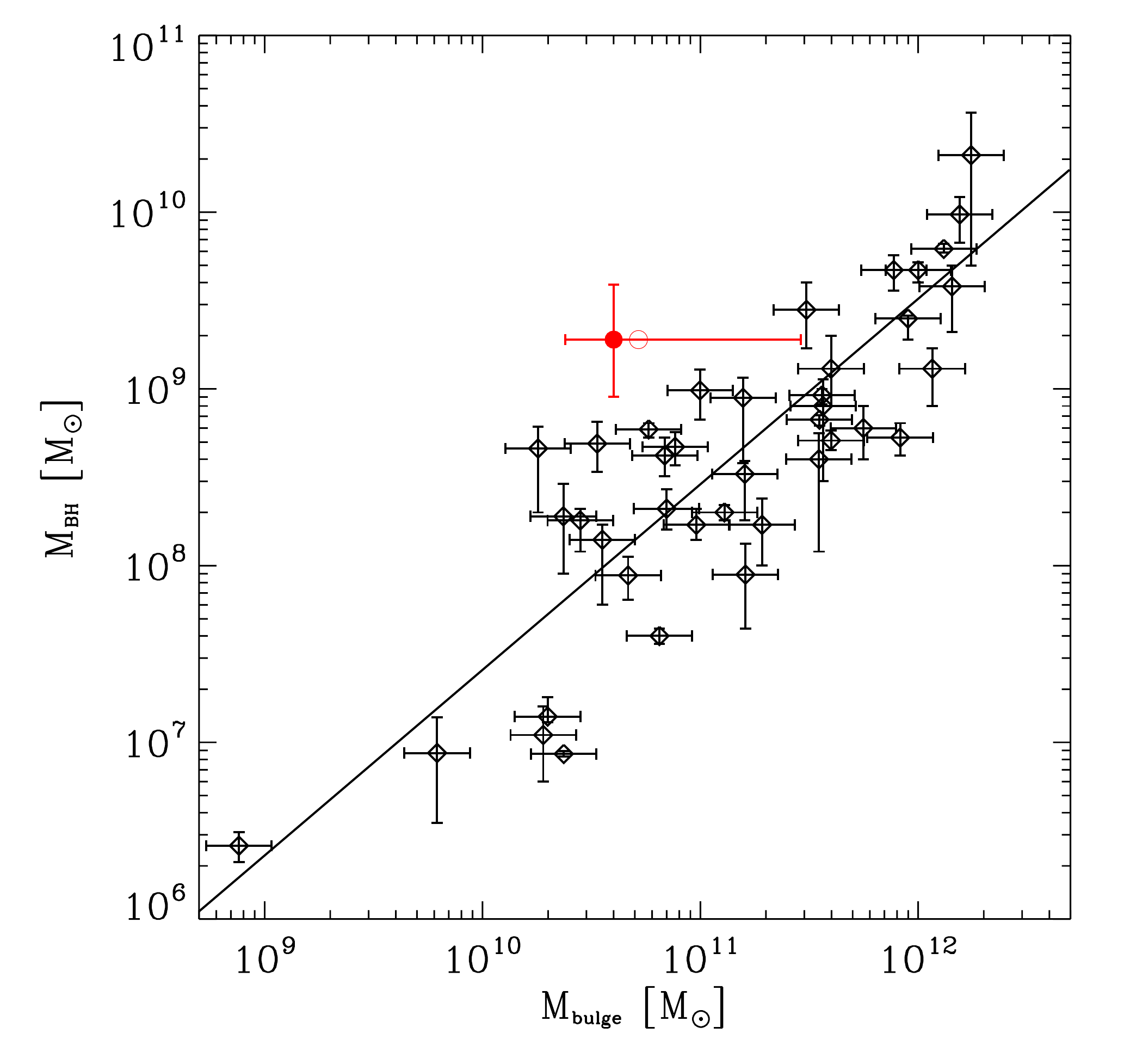} 
   \caption{M$_{\rm BH}$-M$_{\rm bulge}$ relation for  35 early-type galaxies in the local Universe (empty diamonds)  and BR1202 S (red filled circle; the empty filled circle also considers  the contribution from the dispersion-supported mass).  The black line represents the best-fitting relation  estimated by \cite{mcconnell:2013}: $\log(M_{BH}/M_\odot) = 8.46 +1.05\log(M_{bulge}/10^{11} M_\odot)$.
   }
    \label{fig:mbh_mbulge}%
    \end{figure}

Figure \ref{fig:mbh_mbulge} shows the most up-to-date M$_{\rm BH}$-M$_{\rm bulge}$ relation in the local Universe obtained by \cite{mcconnell:2013}. The filled red dot with error bars marks the location of the QSO from the measurements presented above. The empty red circle marks the location of the QSO if the dynamical mass supported by the intrinsic velocity dispersion is taken into account.
We recall that the QSO BH mass is a virial estimate, not from spatially resolved kinematical modeling as  for the black points in the figure. Moreover, the mass of the QSO host galaxy is not the bulge mass, but the total dynamical mass enclosed within 5 kpc. 

The present measurement agrees marginally with the upper envelope of the relation, where the uncertainties on the galaxy mass are taken into account.
The best-fit relation by \cite{mcconnell:2013}, $\log(M_{BH}/M_\odot) = 8.46 +1.05\log(M_{bulge}/10^{11} M_\odot)$, would predict a BH mass of $\sim 1.2\times 10^{8}\msun$, which is  a factor 10 smaller than observed, consistent with previous estimates for luminous quasars at high redshift (see, e.g., \citealt{willott:2013,wang:2013,venemans:2012,merloni:2010,decarli:2010,lamastra:2010,riechers:2009a,peng:2006}).
However, the uncertainties on the disk inclination of the QSO host galaxy are such that this measurement could be consistent with the upper envelope of the local relation and in particular with the most massive local galaxy like BCGs \citep{mcconnell:2012}. 

This is the first dynamical mass estimate of a QSO host galaxy at high redshift  based on the analysis of the spatially resolved kinematics, the only other one is that of a QSO host galaxy at $z\sim 1.3$
\citep{inskip:2011}. All other measurements are either stellar masses based on photometry  \citep{merloni:2010,decarli:2010,peng:2006} or dynamical masses  based on simple virial estimates \citep{willott:2013,wang:2013,venemans:2012,riechers:2009a}. The former are subjected to large uncertainties and selection biases,  which might account for the large 
$M_{BH}/M_{bul}$ observed \citep{lamastra:2010}.
The latter are  subject to large uncertainties  on the unknown geometry and kinematics of the host galaxy. In particular, the present work has shown that it is critical to properly take into account  the disk inclination: if we were to assume an average inclination of 60 degrees, the dynamical masses so obtained would have been inconsistent with the molecular gas masses. 

Therefore, the only possibility to avoid the large uncertainties and selection biases on stellar masses and the large uncertainties on virial mass estimates is to use dynamical host galaxy masses obtained from spatially resolved kinematics. 
Intriguingly, in the only other case in which the dynamical mass of a QSO host galaxy ($z\sim 1.3$) has been determined from a spatially resolved kinematical analysis, the BH mass and the galaxy mass agree with the local relation \citep{inskip:2011}.
This suggests that dynamical mass determinations from submm spectroscopy such as presented here might be much better probes of the cosmological evolution of the $M_{BH}-M_{gal}$ relation. 

\section{Conclusions}\label{sec:conclusions}

We performed a kinematical analysis of the [CII] line emission in the BR 1202-0725 system at $z\sim 4.7$ from ALMA Science Verification observations. The quasar and the submillimeter galaxy are separated by $\sim$24 kpc and are characterized by very high star-formation rates, larger than $\sim 1000\,\mathrm{M_\odot\,yr^{-1}}$.
Our kinematical analysis reveals that these galaxies apparently  have regularly rotating disks, that are seen almost face-on and that indicate dynamical masses of $6\times10^{10}$ \msun, and $4\times10^{10}$ \msun \, for the SMG and QSO host galaxy, respectively. These disks are hotter than the disks of nearby galaxies, since the maximum rotational velocity is similar to the intrinsic velocity dispersion, but they are very similar to the disks of $z\sim 2-3$ Lyman-break galaxies.
If the intrinsic velocity dispersion of the gas disks provides support against gravity, these masses could increase up to  $\sim 9\times10^{10}$ \msun \, and $\sim 5\times10^{10}$ \msun.
Overall, the SMG and the QSO host galaxy are characterized by very similar physical properties in terms of SFRs, molecular gas mass, and dynamical mass.
 
The existence of these rotating disks suggest the the high star formation and black hole accretion rates are not induced by a major-merger event, at variance with the commonly accepted scenario for very massive galaxies and very high SFRs. 
We also detected faint components which, after a spectral deblending, were spatially resolved from the main QSO and SMG emissions. The relative  velocities and positions of these components are consistent with orbital motions within the gravitational potentials generated by the QSO host galaxy and SMG, suggesting that they are smaller galaxies in interaction or gas clouds in accretion flows of tidal streams. Moreover, we did not find any clear spectral evidence for outflows caused by AGN or stellar feedback.

The rotating disks and the spectroscopically detected faint components   suggest that the high SFRs might be induced by interactions or minor mergers with these companions which, however, do not affect the large-scale kinematics of the disks. Alternatively, the strong star formation may be fueled by the accretion of pristine gas from the host halo. Both these processes could explain the relative high intrinsic velocity dispersion.

Finally, the ratio between the black hole mass of the QSO, obtained from new XShooter spectroscopy, and the dynamical mass of the host galaxy might be comparable with the similarly high value found in very massive local galaxies, suggesting that the evolution of black hole galaxy relations is probably better studied with dynamical  than  with stellar host galaxy masses.

New ALMA observations with higher sensitivity and spatial resolution are required to confirm the findings presented here.
\begin{acknowledgements}
We thank the anonymous referee for a very quick and thoughtful report. We thank Leslie Hunt, Sperello di Serego Alighieri, and Claudia Cicone for comments and discussions.
This paper makes use of the following ALMA data: ADS/JAO.ADS/JAO.ALMA\#2011.0.00002.SV. ALMA is a partnership of ESO (representing its member states), NSF (USA) and NINS (Japan), together with NRC (Canada) and NSC and ASIAA (Taiwan), in cooperation with the Republic of Chile. The Joint ALMA Observatory is operated by ESO, AUI/NRAO and NAOJ.
Also based on observations collected at the  European  Southern Observatory  Very  Large Telescope,  Cerro  Paranal,  Chile -- Programs 084.A-0780(B), P.I. P. Molaro.
Support for this publication was provided by the Italian National Institute for Astrophysics (INAF) through PRIN-INAF 2011 "Black hole growth and AGN feedback through the cosmic time" and by the Italian ministry for school, university and reasearch (MIUR) through PRIN-MIUR 2010-2011 "The dark Universe and the cosmic evolution of baryons: from current surveys to Euclid".
\end{acknowledgements}

\end{document}